\PassOptionsToPackage{unicode}{hyperref}
\PassOptionsToPackage{hyphens}{url}
\PassOptionsToPackage{dvipsnames,svgnames,x11names}{xcolor}
\documentclass[
  12pt,
  letterpaper,
]{article}

\usepackage{amsmath,amssymb}
\usepackage{setspace}
\usepackage{iftex}
\ifPDFTeX
  \usepackage[T1]{fontenc}
  \usepackage[utf8]{inputenc}
  \usepackage{textcomp} 
\else 
  \usepackage{unicode-math}
  \defaultfontfeatures{Scale=MatchLowercase}
  \defaultfontfeatures[\rmfamily]{Ligatures=TeX,Scale=1}
\fi
\usepackage{lmodern}
\ifPDFTeX\else  
\fi
\IfFileExists{upquote.sty}{\usepackage{upquote}}{}
\IfFileExists{microtype.sty}{
  \usepackage[]{microtype}
  \UseMicrotypeSet[protrusion]{basicmath} 
}{}
\makeatletter
\@ifundefined{KOMAClassName}{
  \IfFileExists{parskip.sty}{%
    \usepackage{parskip}
  }{
    \setlength{\parindent}{0pt}
    \setlength{\parskip}{6pt plus 2pt minus 1pt}}
}{
  \KOMAoptions{parskip=half}}
\makeatother
\usepackage{xcolor}
\usepackage[margin=1in]{geometry}
\makeatletter
\ifx\paragraph\undefined\else
  \let\oldparagraph\paragraph
  \renewcommand{\paragraph}{
    \@ifstar
      \xxxParagraphStar
      \xxxParagraphNoStar
  }
  \newcommand{\xxxParagraphStar}[1]{\oldparagraph*{#1}\mbox{}}
  \newcommand{\xxxParagraphNoStar}[1]{\oldparagraph{#1}\mbox{}}
\fi
\ifx\subparagraph\undefined\else
  \let\oldsubparagraph\subparagraph
  \renewcommand{\subparagraph}{
    \@ifstar
      \xxxSubParagraphStar
      \xxxSubParagraphNoStar
  }
  \newcommand{\xxxSubParagraphStar}[1]{\oldsubparagraph*{#1}\mbox{}}
  \newcommand{\xxxSubParagraphNoStar}[1]{\oldsubparagraph{#1}\mbox{}}
\fi
\makeatother

\usepackage{longtable,booktabs,array}
\usepackage{calc} 
\usepackage{etoolbox}
\makeatletter
\patchcmd\longtable{\par}{\if@noskipsec\mbox{}\fi\par}{}{}
\makeatother
\IfFileExists{footnotehyper.sty}{\usepackage{footnotehyper}}{\usepackage{footnote}}
\makesavenoteenv{longtable}
\usepackage{graphicx}
\makeatletter
\newsavebox\pandoc@box
\newcommand*\pandocbounded[1]{
  \sbox\pandoc@box{#1}%
  \Gscale@div\@tempa{\textheight}{\dimexpr\ht\pandoc@box+\dp\pandoc@box\relax}%
  \Gscale@div\@tempb{\linewidth}{\wd\pandoc@box}%
  \ifdim\@tempb\p@<\@tempa\p@\let\@tempa\@tempb\fi
  \ifdim\@tempa\p@<\p@\scalebox{\@tempa}{\usebox\pandoc@box}%
  \else\usebox{\pandoc@box}%
  \fi%
}
\def\fps@figure{htbp}
\makeatother
\NewDocumentCommand\citeproctext{}{}
\NewDocumentCommand\citeproc{mm}{%
  \begingroup\def\citeproctext{#2}\cite{#1}\endgroup}
\makeatletter
 \let\@cite@ofmt\@firstofone
 \def\@biblabel#1{}
 \def\@cite#1#2{{#1\if@tempswa , #2\fi}}
\makeatother
\newlength{\cslhangindent}
\newlength{\csllabelwidth}
\newenvironment{CSLReferences}[2] 
 {\begin{list}{}{%
  \setlength{\itemindent}{0pt}
  \setlength{\leftmargin}{0pt}
  \setlength{\parsep}{0pt}
  \ifodd #1
   \setlength{\leftmargin}{\cslhangindent}
   \setlength{\itemindent}{-1\cslhangindent}
  \fi
  \setlength{\itemsep}{#2\baselineskip}}}
 {\end{list}}
\usepackage{calc}

\usepackage{amsmath,amssymb,booktabs,etoolbox,needspace}
\AtBeginEnvironment{longtable}{\small\setstretch{1.0}}
\ifPDFTeX\DeclareUnicodeCharacter{2212}{\ensuremath{-}}\DeclareUnicodeCharacter{0394}{\ensuremath{\Delta}}\fi
\makeatletter
\@ifpackageloaded{caption}{}{\usepackage{caption}}
\AtBeginDocument{%
\ifdefined\contentsname
  \renewcommand*\contentsname{Table of contents}
\else
  \newcommand\contentsname{Table of contents}
\fi
\ifdefined\listfigurename
  \renewcommand*\listfigurename{List of Figures}
\else
  \newcommand\listfigurename{List of Figures}
\fi
\ifdefined\listtablename
  \renewcommand*\listtablename{List of Tables}
\else
  \newcommand\listtablename{List of Tables}
\fi
\ifdefined\figurename
  \renewcommand*\figurename{Figure}
\else
  \newcommand\figurename{Figure}
\fi
\ifdefined\tablename
  \renewcommand*\tablename{Table}
\else
  \newcommand\tablename{Table}
\fi
}
\@ifpackageloaded{float}{}{\usepackage{float}}
\floatstyle{ruled}
\@ifundefined{c@chapter}{\newfloat{codelisting}{h}{lop}}{\newfloat{codelisting}{h}{lop}[chapter]}
\floatname{codelisting}{Listing}

\makeatother
\makeatletter
\@ifpackageloaded{caption}{}{\usepackage{caption}}
\@ifpackageloaded{subcaption}{}{\usepackage{subcaption}}
\makeatother

\usepackage{bookmark}

\IfFileExists{xurl.sty}{\usepackage{xurl}}{} 
\hypersetup{
  pdftitle={Optimal Covariate Adjustment beyond the Average Treatment Effect: Treated-Population and Overlap-Weighted Estimands},
  pdfauthor={Shoki Okubo},
  pdfkeywords={adjustment set, average treatment effect on the
treated, efficiency bound, directed acyclic graph, overlap
weights, semiparametric efficiency},
  colorlinks=true,
  linkcolor={blue},
  filecolor={Maroon},
  citecolor={Blue},
  urlcolor={Blue},
  pdfcreator={LaTeX via pandoc}}

\title{Optimal Covariate Adjustment beyond the Average Treatment Effect:
Treated-Population and Overlap-Weighted
Estimands\thanks{Replication code and data are available at \url{https://github.com/sokubo/paper-estimand-adjustment-replication}. This work was supported by JSPS KAKENHI Grant Number 26K05332.}}
\author{Shoki
Okubo\thanks{Department of Sociology, Toyo University, Tokyo, Japan. Email: okubo080@toyo.jp. Website: sokubo.github.io.}}
\date{September 14, 2026}

\begin{document}
\maketitle
\begin{abstract}
Graphical causal inference supplies a complete theory of efficient
covariate adjustment for the average treatment effect: one adjustment
set, computable from the graph, is optimal under every compatible
distribution. We show that this is a property of the average treatment
effect's inverse-prevalence weights, not of causal estimands in general.
For the average treatment effect on the treated we index the efficiency
bound by the adjustment set and derive exact identities for its change
under treatment-side and outcome-side extensions of a valid set.
Covariates that predict only the treated-arm outcome are exactly
efficiency-neutral, and covariates that predict the control-arm outcome
can strictly increase the bound when the propensity is below one half
--- a reversal of the supplementation lemma whose source is an
arithmetic--geometric-mean inequality that holds for the average
treatment effect and fails for the treated-population estimand. A
construction with two faithful distributions on one graph proves that no
graphical optimality criterion exists for the treated-population
estimand; under no effect modification the ATE-optimal set is
nonetheless optimal among the graphically valid sets, with an exact
expression for its advantage. The results extend to weighted average
treatment effects with propensity-dependent weights, yielding symmetric
thresholds for overlap weights, an estimand-drift phenomenon under
instrument adjustment, and a characterization of constant weights as the
only smooth positive weights for which outcome-side supplementation
never increases the bound. Simulations and the LaLonde data provide
illustrations.
\end{abstract}

\noindent\textbf{Keywords:} adjustment set; average treatment effect on the treated; efficiency bound; directed acyclic graph; overlap weights; semiparametric efficiency

\setstretch{1.5}
\section{Introduction}\label{sec-intro}

Which valid adjustment set should an analyst use? For the population
average treatment effect (ATE), graphical causal inference gives a
complete and strikingly clean answer. Among all valid adjustment sets in
a causal directed acyclic graph (DAG) there is an \emph{optimal} set,
computable from the graph alone, that minimizes the asymptotic variance
under every distribution compatible with the graph: for the ordinary
least squares estimator in linear structural equation models
(\citeproc{ref-henckel2022}{Henckel, Perković, and Maathuis 2022}) and
for the nonparametric efficiency bound and the estimators that attain it
(\citeproc{ref-rotnitzkysmucler2020}{Rotnitzky and Smucler 2020}); Witte
et al. (\citeproc{ref-witte2020}{2020}) characterize the same set as the
parent set of the outcome in a suitable latent projection and connect it
to data-driven variable selection. The optimal set favors parents of the
outcome and shuns instruments; it is the same set for every distribution
compatible with the graph; and the two operations that prove its
optimality --- \emph{supplementing} a valid set with covariates
unrelated to treatment, and \emph{deleting} covariates unrelated to the
outcome --- never increase the variance. The theory has since been
extended to hidden-variable models
(\citeproc{ref-smuclersapienza2022}{Smucler, Sapienza, and Rotnitzky
2022}), to minimal-cost selection
(\citeproc{ref-smuclerrotnitzky2022}{Smucler and Rotnitzky 2022}), to
time-dependent treatments (\citeproc{ref-adenyo2025}{Adenyo et al.
2025}), and, in a first step beyond the ATE, to a weighted controlled
direct effect whose weights range over the marginal distribution of a
mediator (\citeproc{ref-linguogan2025}{Lin, Guo, and Gan 2025}).

Applied studies, however, do not estimate only the ATE. The average
treatment effect on the treated (ATT) is the leading estimand of program
and policy evaluation, of most matching applications, and of the
difference-in-differences tradition; overlap-weighted estimands (ATO)
are an established option under limited overlap
(\citeproc{ref-limorganzaslavsky2018}{Li, Morgan, and Zaslavsky 2018});
and a family of weighted average treatment effects (WATEs) --- with
weights that are known functions of the covariates in Hirano, Imbens,
and Ridder (\citeproc{ref-hiranoimbensridder2003}{2003}) and functions
of the propensity score in Li, Morgan, and Zaslavsky
(\citeproc{ref-limorganzaslavsky2018}{2018}) --- organizes all of these.
These are also the estimands of the \emph{rare-treatment} regime ---
treatment prevalences of a few percent are the norm in policy evaluation
and pharmacoepidemiology --- in which, as we show, the ATE and the ATT
disagree most sharply about what a good adjustment set is.

For the ATT, the efficiency consequences of covariate choice have been
studied from two angles. Hahn (\citeproc{ref-hahn2004}{2004}) showed
that knowing the propensity score to depend on a subset of the
covariates is ancillary for the ATE yet lowers the efficiency bound for
the ATT, and de Luna, Waernbaum, and Richardson
(\citeproc{ref-deluna2011}{2011, sec. 3.3}) used this to compare the ATE
and ATT bounds under maintained covariate-reduction restrictions. White
and Lu (\citeproc{ref-whitelu2011}{2011, secs. II.A--D}), in the same
setting of a valid set \(X_1\) and a valid superset \((X_1, X_2)\) with
\(X_2\) excluded from the propensity score, also compared the asymptotic
variances of the standard efficient ATT estimators that use \(X_1\) only
or \((X_1, X_2)\) without exploiting the exclusion --- Hahn's (1998)
bound evaluated at each set, the quantities this paper calls
\(V_{\mathrm{att}}(S)\) and \(V_{\mathrm{att}}(S \cup W)\) --- recorded
that their difference can have either sign, in contrast to the ATE, and
used causal diagrams to read off which structures make the additional
covariates efficient in Hahn's sense. Kitagawa and Muris
(\citeproc{ref-kitagawamuris2016}{2016}) compared parametric
propensity-score submodels indexed by regressor sets and found the
asymptotic variance of their normalized weighting estimator to decrease
monotonically in the regressor set for the ATE but not for the ATT, an
inflation that disappears under homogeneous effects; and Lu
(\citeproc{ref-lu2015}{2015}) proposed a data-driven criterion that
selects among, or averages over, ATE and ATT estimators based on
different covariate sets --- all valid, or some locally invalid --- by
their estimated asymptotic mean squared error, noting with Hahn
(\citeproc{ref-hahn2004}{2004}) and White and Lu
(\citeproc{ref-whitelu2011}{2011}) that conditioning on more covariates
is not always more efficient. Thus, it is known that adjusting for an
additional outcome predictor can reduce precision when estimating the
ATT.

What this literature does not ask is the question that the graphical ATE
theory answers: given the set of covariates that a causal graph declares
valid, is there a set --- computable from the graph, optimal under every
compatible law --- that an analyst targeting the ATT should adjust for,
and if not, what does the graph still determine? To answer it we
re-express the known comparison in conditional moments --- identity (3)
writes the difference of the two set-indexed bounds in terms of the
propensity and the conditional variances and covariance of the two arm
regressions, for arbitrary valid sets in the unrestricted model, and so
makes the sign of the change and its dependence on the law explicit
(Section 2.3) --- and we then use that expression for what is new here:
the graphical impossibility and rescue theorems and the characterization
across propensity weights.

Our answer differs sharply from that for the ATE. We present five
results.

\emph{First, the adjustment calculus for the ATT} (Theorem 1). We record
the semiparametric variance bound for the ATT at each valid adjustment
set --- Hahn's bound, indexed by the set --- and obtain exact
identities, in conditional moments, for its change when a valid set is
extended on the treatment side (adding covariates unrelated to the
outcome given the set) or on the outcome side (adding covariates
unrelated to treatment given the set). The treatment-side identity
mirrors the ATE: instruments are pure variance loads. The outcome-side
identity does not. Adding a covariate that predicts only the
\emph{treated} arm --- strictly beneficial for the ATE --- is exactly
efficiency-neutral for the ATT (Corollary 1); and adding a covariate
that predicts the \emph{control} arm can strictly \emph{increase} the
ATT bound when treatment is rare and effect modification is
arm-asymmetric (Corollary 2), reversing the supplementation lemma of
Rotnitzky and Smucler (\citeproc{ref-rotnitzkysmucler2020}{2020}). The
reversal has an elementary source: the ATE bound weights the two arms by
\(1/e\) and \(1/(1-e)\), and an arithmetic--geometric-mean inequality in
these weights protects supplementation (Corollary 4); the ATT weights,
\(e\) and \(e^{2}/(1-e)\), break the balance.

\emph{Second, ATT-optimality is not graphical} (Theorem 2). We construct
a DAG and two strictly positive distributions, both Markov and faithful
to it, whose ATT-optimal valid adjustment sets differ. No map from
graphs to adjustment sets can therefore be ATT-optimal for all
compatible distributions, in contrast to the ATE. The same construction
proves that the ATE-optimal set can be strictly ATT-suboptimal, with a
closed-form variance penalty.

\emph{Third, when it is} (Theorem 3). Under no effect modification --- a
sufficient condition for the ATT and the ATE to be numerically equal ---
the ATE-optimal set \(O(G)\) is ATT-optimal as well among the
graphically valid sets, with an exact two-term expression for its
advantage over any such competitor; the restriction to graphically valid
competitors is necessary, and Section 6 gives a faithful example in
which a set valid only at the law beats \(O(G)\). The proof follows the
supplement-then-delete architecture of Rotnitzky and Smucler
(\citeproc{ref-rotnitzkysmucler2020}{2020}) with the ATT increments of
Theorem 1 in place of the ATE increments; the family of Theorem 2 shows
the condition cannot be dropped.

\emph{Fourth, the weighted class} (Lemma 2, Theorem 4, Corollaries 6--7,
Proposition 1). For every WATE whose weight is a smooth function \(h\)
of the propensity score we record the adjustment-set-indexed bound ---
its influence function, with the derivative term that an estimated
propensity contributes, is that of Wang, Liu, and Yang
(\citeproc{ref-wang2025}{2025}) --- and extend the outcome-side calculus
to it, recovering the ATE (\(h \equiv 1\)) and ATT (\(h(e) = e\)) as
instances. For the overlap weights \(h(e) = e(1-e)\), we obtain a
symmetric pair of pointwise thresholds: a control-arm-only predictor
harms ATO efficiency where \(e < 1/3\), a treated-arm-only predictor
where \(e > 2/3\) (Corollary 6); and we show that constant weights are
the only smooth positive weights for which outcome-side supplementation
never increases the bound, whatever the joint predictive structure of
the two arms (Corollary 7). We also isolate a phenomenon with no ATE or
ATT analogue: for overlap-type weights, adjusting for an instrument can
\emph{change the estimand}, not merely the variance (Proposition 1), so
that ``deleting instruments'' is not even a well-posed efficiency
comparison for the ATO until a canonical weight is fixed.

\emph{Fifth, finite samples and data.} A simulation study confirms that,
in the designs simulated, the bound orderings govern finite-sample root
mean squared error and the influence-function standard errors are
calibrated. A random-structure scan shows that the phenomenon of Theorem
2 is not knife-edge (in 10 of 49 random four-covariate structures with
arm-asymmetric modification the ATE- and ATT-optimal graphically valid
sets differed, with exact ATT penalties of 2.9 to 25.9 percent for using
the ATE-optimal set; the scan counts a divergence when that penalty
exceeds 1 percent, and the frequency is relative to the generating
distribution stated in Appendix B). A sensitivity illustration on the
canonical ATT data of LaLonde, in the Dehejia--Wahba subsample --- 185
trainees against 15,992 Current Population Survey controls, a mixing
proportion of 1.1 percent --- does not estimate the sign functional of
Theorem 1 but shows the setting it describes: the seven leave-one-out
specifications that leave the estimate essentially unchanged have
standard errors at most 1.6 percent below the full model's. For each
single-covariate deletion, the mean squared change in the fitted control
predictions on the treated units amounts to at most a few percent of
their residual earnings variance. Substantive validity arguments, not
precision, should govern the choice of set.

Section 2 sets up the problem and the class of estimands; Section 3
derives the bounds and records their attainment; Sections 4--6 develop
the ATT theory; Section 7 treats the weighted class; Sections 8--9
report simulations and the application; Section 10 extends the results
to hidden variables; Section 11 discusses practice. Proofs are in
Appendix A, numerical verification in Appendix B.

\section{Setup}\label{sec-setup}

\subsection{Causal model and valid adjustment
sets}\label{causal-model-and-valid-adjustment-sets}

Let \(Z = (Z_1, \dots, Z_J)\) be pretreatment covariates,
\(A \in \{0,1\}\) a treatment, and \(Y\) an outcome with potential
outcomes \((Y(0), Y(1))\). Assume consistency,
\(Y = A Y(1) + (1-A) Y(0)\). For a subvector \(S \subseteq Z\) write
\(e_S = P(A=1 \mid S)\), \(\mu_a(S) = E(Y \mid A=a, S)\),
\(\sigma_a^2(S) = \mathrm{var}(Y \mid A=a, S)\),
\(\tau_S = \mu_1(S) - \mu_0(S)\), and \(p = P(A=1) \in (0,1)\).

\textbf{Definition 1 (valid adjustment set).} \(S \subseteq Z\) is
\emph{valid} at the law \(P\) if (i)
\((Y(0), Y(1)) \perp\!\!\!\perp A \mid S\), and (ii) there is \(c > 0\)
with \(c \le e_S \le 1 - c\) almost surely. We write \(\mathcal{A}(P)\)
for the class of sets valid at \(P\).

When a causal DAG \(G\) over \((Z, A, Y)\) is maintained, in the sense
that the observed law is Markov to \(G\) and the potential outcomes are
generated by the nonparametric structural equation model with
independent errors that \(G\) represents, (i) holds for every \(S\)
satisfying the adjustment criterion of Shpitser, VanderWeele, and Robins
(\citeproc{ref-shpitser2010}{2010}) --- in the generalized form of
Perković et al. (\citeproc{ref-perkovic2018}{2018}) when only an
equivalence class of graphs is known --- and it holds for every
distribution in the model. We write \(\mathcal{A}(G)\) for the class of
sets satisfying the adjustment criterion relative to \(G\)
(\emph{graphically valid} sets) and assume positivity for each of them.
The two classes differ in an important way: \(\mathcal{A}(G) \subseteq
\mathcal{A}(P)\) for every \(P\) in the model, but the inclusion can be
strict, because a particular law may satisfy conditional independences
between the potential outcomes and \(A\) that the graph does not encode,
and faithfulness of the observed law to \(G\) does not exclude them. The
graph therefore supplies a conservative guarantee of validity, not an
enumeration of the sets valid at a given law. Results below that compare
two named sets under explicitly stated conditional independences
(Theorems 1, 2 and 4, Proposition 1) are stated on \(\mathcal{A}(P)\);
results that import graphical facts about \(O(G)\) (Lemma 4, Theorem 3)
are stated on \(\mathcal{A}(G)\), as are the corresponding ATE results
of Rotnitzky and Smucler (\citeproc{ref-rotnitzkysmucler2020}{2020}),
whose Definition 2 requires validity under every distribution in the
graphical model. Theorem 5 is stated on a class of sets whose defining
premises are such graphical facts. Throughout we assume
\(E\{Y(a)^2\} < \infty\).

\subsection{Estimands}\label{estimands}

The treated-population estimand is \[
\psi \;=\; E\{Y(1) - Y(0) \mid A = 1\}
 \;=\; \frac{E\{e_S\,\tau_S\}}{E\{e_S\}}
 \qquad \text{for any valid } S.
\] More generally, for a weight function \(h : (0,1) \to [0, \infty)\)
that is continuously differentiable, the \emph{weighted average
treatment effect} indexed by \(h\) and by a valid set \(S\) is \[
\psi_h(S) \;=\; \frac{E\{h(e_S)\,\tau_S\}}{E\{h(e_S)\}}.
\] The ATE is \(h \equiv 1\), the ATT is \(h(e) = e\), the average
treatment effect on the controls (ATC) is \(h(e) = 1 - e\), and the
overlap-weighted ATO of Li, Morgan, and Zaslavsky
(\citeproc{ref-limorganzaslavsky2018}{2018}) is \(h(e) = e(1-e)\); the
class extends the weighted average treatment effects of Hirano, Imbens,
and Ridder (\citeproc{ref-hiranoimbensridder2003}{2003}), whose weights
are known functions of the covariates, to weights that depend on the law
through the propensity score, as in Li, Morgan, and Zaslavsky
(\citeproc{ref-limorganzaslavsky2018}{2018}) and Wang, Liu, and Yang
(\citeproc{ref-wang2025}{2025}). Two facts about the index \(S\) are
central. When \(h\) is affine, \(h(e) = a + b e\) --- the ATE, the ATT,
the ATC and their mixtures --- \(\psi_h(S)\) does not depend on which
valid \(S\) is used: by validity
\(E\{\tau_S\} = \psi_{\mathrm{ate}} = E\{Y(1) - Y(0)\}\) and
\(E\{e_S \tau_S\} = E\{A\, \tau_S\} = E[A\{Y(1)-Y(0)\}] = p\,\psi\), so
that \[
\psi_h(S) \;=\; \frac{a\,\psi_{\mathrm{ate}} + b\,p\,\psi}{a + b\,p}
\qquad \text{for every valid } S.
\] For any nonaffine weight the estimand is
\emph{adjustment-set-indexed}: \(\psi_h(S)\) and \(\psi_h(S')\) are
generally different numbers for two valid sets (Proposition 1 quantifies
this). Efficiency comparisons across sets are therefore well posed
without further conventions exactly for the affine class, of which the
ATE and the ATT are the generators. This is one reason the ATT is the
natural first case.

\subsection{The comparison framework}\label{the-comparison-framework}

We compare bounds across valid sets in the sense of Rotnitzky and
Smucler (\citeproc{ref-rotnitzkysmucler2020}{2020}): for each valid
\(S\), \(\chi_S(P) = E\{\mu_1(S) - \mu_0(S) \mid A = 1\}\) is a
functional of the observed law of \((S, A, Y)\) that equals \(\psi\)
under Definition 1; the quantity compared is the nonparametric
efficiency bound for \(\chi_S\) based on data \((S, A, Y)\), denoted
\(V_{\mathrm{att}}(S)\), and analogously \(V_{\mathrm{ate}}(S)\) and
\(V_h(S)\). Any estimator that is regular and asymptotically linear in
the nonparametric model for \((S, A, Y)\) necessarily has influence
function \(\varphi_S\) (Lemma 1 below) and attains the bound. Augmented
inverse probability weighting with cross-fitted nuisances and targeted
maximum likelihood are members of this class under the conditions of
Section 3.1. The ordering of bounds is therefore the ordering of the
large-sample variances of the estimators in this class.

\emph{Remark 1 (the valid class is itself estimand-dependent).} For the
ATT, (i) of Definition 1 can be weakened to
\(Y(0) \perp\!\!\!\perp A \mid S\) and (ii) to \(e_S \le 1 - c\): the
treated arm is observed directly, so only the control-arm counterfactual
requires adjustment (\citeproc{ref-deluna2011}{de Luna, Waernbaum, and
Richardson 2011}). The class of ATT-valid sets is therefore a superset
of the ATE-valid class. To keep comparisons conservative, our results
are stated on the common (ATE-valid) class. Enlarging the class does not
by itself strengthen a nonexistence statement, since a newly admitted
set could dominate the old ones; for the construction of Theorem 2 the
two classes coincide, because every set that is ATT-valid there contains
the confounder \(Z\) and is therefore also ATE-valid.

\emph{Remark 2 (what the comparison does and does not say).}
\(V_{\mathrm{att}}(S)\) and \(V_{\mathrm{att}}(S \cup W)\) are the
bounds for two different functionals of the observed law, \(\chi_S\) and
\(\chi_{S \cup W}\), each in a model that places no restriction on the
law of \((S, W, A, Y)\). The two functionals coincide with \(\psi\) when
both sets are valid, and with each other on the observed-law submodel
\(W \perp\!\!\!\perp A \mid S\); outside that submodel they are in
general different parameters, so nothing prevents
\(V_{\mathrm{att}}(S \cup W) > V_{\mathrm{att}}(S)\); Section 4 shows
that it happens. It does not follow that observing \(W\) can be harmful.
If \(W \perp\!\!\!\perp A \mid S\) were maintained as a restriction on
the model, both \(\varphi_S\) and \(\varphi_{S \cup W}\) would be
gradients of the single parameter \(\psi\) in the restricted model,
whose efficiency bound is therefore at most
\(\min\{V_{\mathrm{att}}(S), V_{\mathrm{att}}(S \cup W)\}\); the
efficient estimator of that model, which would exploit the restriction,
is not an adjustment estimator of the kind compared here (the efficiency
implications of such functional restrictions on the propensity score are
the subject of Hahn (\citeproc{ref-hahn2004}{2004})). As in the ATE
theory of Rotnitzky and Smucler
(\citeproc{ref-rotnitzkysmucler2020}{2020}), we order the estimators
that adjust for \(S\) or for \(S \cup W\) without using such
restrictions. ``Adjusting for \(W\) without the restriction can lose''
is the correct reading of the results below; ``observing \(W\) loses''
is not.

\emph{Relation to earlier ATT efficiency comparisons.} Three earlier
comparisons of ATT precision across covariate choices provide context
for the present comparison. In Hahn (\citeproc{ref-hahn2004}{2004}) and
de Luna, Waernbaum, and Richardson (\citeproc{ref-deluna2011}{2011, sec.
3.3}) the comparison is between the bound of the unrestricted model and
the bound of a model in which a covariate-reduction restriction --- on
the propensity score, or on the outcome regressions --- is known and
maintained: the ATT bound falls under a propensity restriction while the
ATE bound does not. That is a comparison of models for a single
functional, and Remark 2 explains why it is not the comparison made
here.

White and Lu (\citeproc{ref-whitelu2011}{2011, sec. II.A}) display, for
a valid set \(X_1\) and a valid superset \((X_1, X_2)\), the asymptotic
variances of the standard efficient ATT estimators that use \(X_1\) only
or \((X_1, X_2)\) without exploiting any restriction --- Hahn's (1998)
bound evaluated at each set, in our notation \(V_{\mathrm{att}}(S)\) and
\(V_{\mathrm{att}}(S \cup W)\) --- alongside Hahn's restricted-model
bound, and state in their §II.D that when \(X_2\) is excluded from the
propensity score the sign of the difference can be positive or negative,
whereas for the ATE the larger set always does at least as well. That is
exactly the comparison of Theorem 1(b), and the identity (3) below is
the closed form of their difference: subtracting the two displayed
bounds and applying the conditional variance decomposition under
\(W \perp\!\!\!\perp A \mid S\) gives (3) (Appendix A.2 does this
through the influence functions). What (3) adds is not the comparison
but its anatomy --- which conditional moments of the law,
\((e_S, V_0, C_{10})\), decide the sign, and why the ATE's weights
protect it --- and it is this anatomy that the graphical results of
Sections 5--7 use.

In Kitagawa and Muris (\citeproc{ref-kitagawamuris2016}{2016}) the
objects compared are normalized inverse-odds weighting estimators of the
ATT built on parametric propensity models with different regressor sets;
their asymptotic variance is nonmonotone in the regressor set for the
ATT and monotone for the ATE, and the nonmonotonicity disappears under
homogeneous effects. When the covariates are discrete and the propensity
model is saturated within each regressor set, their estimator reduces to
the cell-wise estimator \(\bar Y_1 - \sum_s (n_{1s}/n_1)\,\bar Y_{0s}\),
which is also the estimator that attains \(V_{\mathrm{att}}(S)\) in the
discrete case (Section 8). In this setting, their comparison across
regressor sets is again a comparison of the bounds studied here, and (3)
agrees with the finite-cell specialization of their influence-function
expansion.

\section{The adjustment-set-indexed bounds}\label{sec-bounds}

\textbf{Lemma 1 (ATT bound).} Let \(S\) be valid. In the nonparametric
model for \((S, A, Y)\), the functional \(\chi_S\) is pathwise
differentiable with efficient influence function \[
\varphi_S \;=\; \frac{1}{p}\Bigl[ A\{Y - \mu_0(S) - \psi\}
 \;-\; \frac{e_S}{1-e_S}\,(1-A)\{Y - \mu_0(S)\} \Bigr],
\] and efficiency bound \[
V_{\mathrm{att}}(S) \;=\; \frac{1}{p^{2}}\,
E\Bigl[\, e_S\,\sigma_1^2(S) \;+\; \frac{e_S^{2}}{1-e_S}\,\sigma_0^2(S)
 \;+\; e_S\,(\tau_S - \psi)^{2} \Bigr].
\tag{1}
\]

The full-covariate case recovers the bound of Hahn
(\citeproc{ref-hahn1998}{1998}). The bound has three components --- the
treated-arm variance, the control-arm variance, and the dispersion of
the conditional effect --- and the analysis below focuses on how
different estimands weight them. Lemma 2 gives the general result.

\textbf{Lemma 2 (WATE bound).} Let \(S\) be valid and \(h\) continuously
differentiable on \([c, 1-c]\) with \(D_S = E\{h(e_S)\} > 0\). The
functional \(P \mapsto \psi_h(S)\) is pathwise differentiable in the
nonparametric model for \((S, A, Y)\) with efficient influence function
\[
\varphi_{h,S} = \frac{1}{D_S}\Bigl[ h(e_S)\{\tau_S - \psi_h\}
 + h'(e_S)\{\tau_S - \psi_h\}(A - e_S)
 + h(e_S)\Bigl\{ \frac{A(Y-\mu_1)}{e_S} - \frac{(1-A)(Y-\mu_0)}{1-e_S} \Bigr\} \Bigr]
\] and efficiency bound \[
V_h(S) = \frac{1}{D_S^{2}}\,
E\Bigl[\, \frac{h(e_S)^{2}}{e_S}\,\sigma_1^2(S)
 + \frac{h(e_S)^{2}}{1-e_S}\,\sigma_0^2(S)
 + (\tau_S - \psi_h)^{2}\,\bigl\{ h(e_S)^{2} + h'(e_S)^{2}\, e_S (1-e_S) \bigr\} \Bigr].
\tag{2}
\]

Lemma 2 restates the influence function given in Theorem 1 of Wang, Liu,
and Yang (\citeproc{ref-wang2025}{2025}) (Theorem 3.1 of the arXiv
version). These authors treat the propensity as unknown and note that
with a known propensity the derivative term disappears, which is the
general form of the non-ancillarity of the propensity score for the ATT
established by Hahn (\citeproc{ref-hahn1998}{1998}, Theorems 1--2):
knowledge of the propensity lowers the ATT bound by exactly this term
while leaving the ATE bound unchanged. The printed pseudo-outcome of
Wang, Liu, and Yang (\citeproc{ref-wang2025}{2025}) carries the
conditional effect with the opposite sign, which their own derivations
and our Appendix A.5 show to be a misprint. We index the bound by the
adjustment set because Theorem 4 compares the bounds across sets.
Setting \(h \equiv 1\) gives the ATE bound
\(V_{\mathrm{ate}}(S) = E[\sigma_1^2/e_S + \sigma_0^2/(1-e_S) + (\tau_S - \psi_{\mathrm{ate}})^2]\);
setting \(h(e) = e\) gives \(h^2/e = e\), \(h^2/(1-e) = e^2/(1-e)\), and
\(h^2 + h'^2 e(1-e) = e^2 + e(1-e) = e\), which is (1). The derivative
term \(h'(e)^2 e(1-e)\) is the price of estimating a
propensity-dependent weight; it vanishes for the ATE and is absorbed for
the ATT. Table 1 collects the three weights.

\begin{longtable}[]{@{}
  >{\raggedright\arraybackslash}p{(\linewidth - 8\tabcolsep) * \real{0.1600}}
  >{\raggedright\arraybackslash}p{(\linewidth - 8\tabcolsep) * \real{0.1600}}
  >{\raggedright\arraybackslash}p{(\linewidth - 8\tabcolsep) * \real{0.2200}}
  >{\raggedright\arraybackslash}p{(\linewidth - 8\tabcolsep) * \real{0.2200}}
  >{\raggedright\arraybackslash}p{(\linewidth - 8\tabcolsep) * \real{0.2400}}@{}}
\caption{Weights on the three components of the bound (2), up to the
normalization \(D_S^{-2}\).}\label{tbl-weights}\tabularnewline
\toprule\noalign{}
\begin{minipage}[b]{\linewidth}\raggedright
Estimand
\end{minipage} & \begin{minipage}[b]{\linewidth}\raggedright
\(h(e)\)
\end{minipage} & \begin{minipage}[b]{\linewidth}\raggedright
weight on \(\sigma_1^2\)
\end{minipage} & \begin{minipage}[b]{\linewidth}\raggedright
weight on \(\sigma_0^2\)
\end{minipage} & \begin{minipage}[b]{\linewidth}\raggedright
weight on \((\tau-\psi)^2\)
\end{minipage} \\
\midrule\noalign{}
\endfirsthead
\toprule\noalign{}
\begin{minipage}[b]{\linewidth}\raggedright
Estimand
\end{minipage} & \begin{minipage}[b]{\linewidth}\raggedright
\(h(e)\)
\end{minipage} & \begin{minipage}[b]{\linewidth}\raggedright
weight on \(\sigma_1^2\)
\end{minipage} & \begin{minipage}[b]{\linewidth}\raggedright
weight on \(\sigma_0^2\)
\end{minipage} & \begin{minipage}[b]{\linewidth}\raggedright
weight on \((\tau-\psi)^2\)
\end{minipage} \\
\midrule\noalign{}
\endhead
\bottomrule\noalign{}
\endlastfoot
ATE & \(1\) & \(1/e\) & \(1/(1-e)\) & \(1\) \\
ATT & \(e\) & \(e\) & \(e^{2}/(1-e)\) & \(e\) \\
ATO & \(e(1-e)\) & \(e(1-e)^{2}\) & \(e^{2}(1-e)\) &
\(e(1-e)\{1 - 3e + 3e^{2}\}\) \\
\end{longtable}

Two structural readings. The ATE weights the arms \emph{inversely} to
their prevalence --- the arm that is rare is the arm whose variance
matters --- while the ATT weights them \emph{directly} by prevalence in
the treated population, so that the control-arm weight \(e^2/(1-e)\) is
smaller than the treated-arm weight \(e\) by the factor \(e/(1-e)\) ---
an additional factor of \(e\) in the small-\(e\) expansion; the
effect-dispersion term, however, retains weight \(e\), which is what
drives Section 4. And the ATO weights are symmetric under
\(e \mapsto 1 - e\) with the arms exchanged, which will produce a
symmetric pair of thresholds in Section 7.

\subsection{Attainment}\label{attainment}

The bounds are attained. For the ATT, the augmented inverse probability
weighted estimator built on \(\varphi_S\) with cross-fitted nuisance
estimates of \((e_S, \mu_0(S))\) is regular and asymptotically linear
with influence function \(\varphi_S\) under the conditions of
Chernozhukov et al. (\citeproc{ref-chernozhukov2018}{2018}, Assumption
5.1 and Theorem 5.1): overlap, bounded moments, consistency of each
nuisance estimator in \(L_2(P)\), and the product rate
\(\|\hat e_S - e_S\|_{P,2}\,\|\hat\mu_0 - \mu_0\|_{P,2} = o_P(n^{-1/2})\);
targeted maximum likelihood estimators attain the same bound under
analogous conditions (\citeproc{ref-vanderlaanrose2011}{van der Laan and
Rose 2011, chap. 8}). For a WATE with a nonaffine propensity-dependent
weight, the requirement is stronger because the derivative term
\(h'(e_S)\) in \(\varphi_{h,S}\) introduces into the one-step
estimator's remainder a pure propensity-error term that outcome modeling
does not cancel: with \(q = \hat e_S\), the population mean of the
estimating function at the true target contains
\(E[\{h(q) + h'(q)(e_S - q) - h(e_S)\}(\tau_S - \psi_h)]\), which for
the overlap weight equals \(E[(q - e_S)^2(\tau_S - \psi_h)]\) exactly.
The term vanishes identically for affine \(h\) and is generally present
for nonaffine \(h\) (it can vanish at particular laws, for instance when
\(\tau_S\) is constant). It is bounded by
\(|E[(q - e_S)^2(\tau_S - \psi_h)]| \le \|\tau_S - \psi_h\|_\infty
\|q - e_S\|_{P,2}^{2}\); a bounded conditional effect therefore gives a
convenient sufficient condition under which the ordinary
propensity-error rate controls the curvature remainder.

\emph{A sufficient regime.} Asymptotic linearity of the cross-fitted
one-step estimator with the overlap weight holds under the following
conditions: consistency of the nuisance estimators; overlap for both the
true and estimated propensity scores; the nondegeneracy condition
\(D_S > 0\) and the moment conditions of the ATT case; a bounded
conditional effect \(\sup_s |\tau_S(s)| < \infty\); the product rates
for the nuisance pairs \((e_S, \mu_a(S))\); and the propensity rate
\(\|\hat e_S - e_S\|_{P,2} = o_P(n^{-1/4})\). This regime is sufficient,
not necessary: the curvature term can vanish at particular laws, and the
remainder can also be controlled directly, since
\(|E[(q - e_S)^2(\tau_S - \psi_h)]| \le \|\tau_S - \psi_h\|_{P,2}
\|q - e_S\|_{P,4}^{2}\), so that for a square-integrable conditional
effect the rate \(\|\hat e_S - e_S\|_{P,4} = o_P(n^{-1/4})\) suffices to
control this remainder without boundedness.

\emph{Two examples.} Neither the propensity rate nor the effect bound in
the simple sufficient regime may be omitted without replacement
restrictions; the examples do not establish necessary conditions for
asymptotic linearity at every law. A propensity estimated at the rate
\(n^{-1/5}\) with exactly known outcome regressions can produce a bias
of order \(n^{-2/5}\). With an unbounded conditional effect, a
propensity error concentrated where the effect is large defeats the
\(L_2\) quarter rate --- with \(S \sim U(0, 1)\), \(e_S \equiv 1/2\),
\(\tau_S = S^{-1/5}\) (finite fourth moments, unit residual variance),
known outcome regressions and
\(\hat e_S = 1/2 + 0.1\,\mathbf{1}\{S < n^{-11/20}\}\), the propensity
error is \(0.1\, n^{-11/40} = o(n^{-1/4})\) in \(L_2\) (its \(L_4\)
error, \(0.1\, n^{-11/80}\), is not) yet the population solution of the
estimating equation is biased by \(0.05\, n^{-11/25}\{1 + o(1)\}\),
which is not \(o(n^{-1/2})\).

\emph{Source conditions.} Wang, Liu, and Yang
(\citeproc{ref-wang2025}{2025}) give the corresponding conditions: their
Condition 1 (p.~7), for estimators without sample splitting, bounds the
product
\(\|\hat e - e\|_{2}\{\sum_a \|\hat\mu_a - \mu_a\|_2 + \|h'(\hat e) - h'(e)\|_2\}\)
(in our notation) by \(o_P(n^{-1/2})\), and their Condition 6 with
Remark 3 (p.~10), for the cross-fitted estimator, replaces the
derivative-error term by a supremum over the path from \(e\) to
\(\hat e\) involving the second and third derivatives of the weight,
which for the overlap weight reduces to the squared propensity error;
the bound on the conditional effect is imposed in their Condition 4
(p.~7, which also bounds \(h(\hat e)/\hat e\) and
\(h(\hat e)/(1 - \hat e)\)) and Condition 10 (p.~10), and the rate
conditions alone are not their full set of assumptions.

\emph{Attainment in the discrete designs.} In the discrete designs of
Section 8 and Appendix B, the large-sample variance of the saturated
cell-wise plug-ins equals the bound, allowing the bounds to be checked
by simulation. At finite \(n\), the Monte Carlo variance differs from
the bound by at most 11 percent: it is 6 to 11 percent above the bound
under rare treatment at \(n = 500\), and its relative deviation from the
bound ranges from \(-6\) to \(+9\) percent in the other cells (Section
8, Appendix B). Because every estimator in this class has the same
limiting variance for a given \(S\), the ordering of bounds across sets
is the ordering of the large-sample variances of the estimators in the
class. Estimators with parametric nuisances, such as those of Section 9,
lie outside this class unless the models are correct.

\section{The ATT adjustment calculus}\label{sec-calculus}

Throughout this section, \(S\) and \(S \cup W\) are both valid. Let
\(\kappa(x) = x^2/(1-x)\) be the control-arm weight in (1); this
function is strictly convex on \([0, 1)\). For the outcome-side result
write, conditionally on \(S\), \[
V_a(S) = \mathrm{var}\{\mu_a(S \cup W) \mid S\}, \qquad
C_{10}(S) = \mathrm{cov}\{\mu_1(S \cup W),\, \mu_0(S \cup W) \mid S\}.
\] (\(V_a\) with an arm subscript is a conditional variance;
\(V_{\mathrm{att}}\), \(V_{\mathrm{ate}}\) and \(V_h\) are bounds.)

\textbf{Theorem 1 (deletion and supplementation for the ATT).}

\textbf{(a) Treatment-side extension.} If
\(Y \perp\!\!\!\perp W \mid A, S\) --- in a causal DAG, whenever \(Y\)
and \(W\) are \(d\)-separated given \(\{A\} \cup S\) --- then \[
V_{\mathrm{att}}(S \cup W) - V_{\mathrm{att}}(S)
 \;=\; \frac{1}{p^{2}}\,
 E\Bigl[ \sigma_0^2(S)\,\bigl\{ E\bigl(\kappa(e_{S\cup W}) \bigm| S\bigr)
  - \kappa(e_S) \bigr\} \Bigr] \;\ge\; 0,
\] with equality if and only if \(e_{S \cup W} = e_S\) almost surely on
\(\{\sigma_0^2(S) > 0\}\).

\textbf{(b) Outcome-side extension.} If \(W \perp\!\!\!\perp A \mid S\),
then \[
V_{\mathrm{att}}(S \cup W) - V_{\mathrm{att}}(S)
 \;=\; \frac{1}{p^{2}}\,
 E\Bigl[\, e_S \Bigl\{ \frac{1 - 2 e_S}{1 - e_S}\, V_0(S)
  \;-\; 2\, C_{10}(S) \Bigr\} \Bigr].
\tag{3}
\]

\emph{Proof.} Appendix A.2. Both parts are exact identities; (a) then
follows from conditional Jensen applied to \(\kappa\). \(\square\)

Part (a) is the ATT analogue of the deletion lemma of Rotnitzky and
Smucler (\citeproc{ref-rotnitzkysmucler2020}{2020}, Lemma 5):
instruments are pure variance loads for the ATT exactly as for the ATE,
and the entire loss is carried by the control-arm term. (Their
bias-amplifying role under residual confounding
(\citeproc{ref-ding2017}{Ding, VanderWeele, and Robins 2017}) lies
outside Definition 1, which fixes validity and asks only about
variance.) Part (b) has no ATE analogue, and the remainder of this
section unpacks it.

\textbf{Corollary 1 (treated-arm predictors are efficiency-neutral).} In
setting (b), if \(\mu_0(S \cup W) = \mu_0(S)\) almost surely --- \(W\)
predicts the outcome only in the treated arm, given \(S\) --- then
\(V_0(S) = C_{10}(S) = 0\) and
\(V_{\mathrm{att}}(S \cup W) = V_{\mathrm{att}}(S)\) exactly. For the
ATE the same extension gives
\(V_{\mathrm{ate}}(S\cup W)-V_{\mathrm{ate}}(S)
= -E[V_1(S)(1-e_S)/e_S] < 0\) whenever \(V_1(S)\) is nondegenerate.

The treated arm of the ATT is averaged directly, with weight \(A/p\) and
no reweighting, so nothing is gained by predicting it. The covariates
that matter for ATT efficiency are the predictors of the
\emph{control-arm} outcome --- a distinction no DAG over the observed
variables records, since both kinds are parents of \(Y\); de Luna,
Waernbaum, and Richardson (\citeproc{ref-deluna2011}{2011, sec. 4.1})
make the same observation about translating separate graphs for \(Y(0)\)
and \(Y(1)\) into one graph for the observed outcome.

\textbf{Corollary 2 (supplementation reversal).} In setting (b), if
\(C_{10}(S) \le 0\) and \(e_S \le 1/2\) almost surely, then
\(V_{\mathrm{att}}(S \cup W) \ge V_{\mathrm{att}}(S)\), strictly if the
event \(\{V_0(S) > 0,\; e_S < 1/2\}\) has positive probability: adding a
genuine outcome predictor strictly \emph{harms} ATT efficiency. If
\(C_{10}(S) \ge 0\) and \(e_S \ge 1/2\) almost surely, the extension
helps.

The mechanism deserves a sentence. Adding \(W\) removes
\(\mathrm{var}\{\mu_0(S\cup W) \mid S\}\) from the control-arm residual
variance, a gain the ATT weights by \(e^2/(1-e)\); but it also moves
that same variation into the dispersion of the conditional effect
\(\tau_{S \cup W}\), a loss the ATT weights by \(e\). When treatment is
rare the loss outweighs the gain. For the ATE the gain is weighted by
\(1/(1-e) \ge 1\) against a loss weighted by \(1\), and the loss can
never win. For the ATC the picture is the mirror image under
\(A \mapsto 1-A\): control-arm predictors are efficiency-neutral, and
treated-arm predictors can harm when \(e_S > 1/2\) (the arm-specific
coefficients of (6) below at \(h(e) = 1-e\) are \((1-e)(2e-1)/e\) and
\(0\)).

\emph{Remark 3 (the rare-treatment limit).} Rarity does not make the
outcome-side increment negligible relative to the bound: along a path on
which treatment becomes rare, the relative increment, where positive,
increases monotonically to a nonzero limit whenever the leading term of
the bound, \(E[\tilde e_S\{\sigma_1^2(S) + (\tau_S - \psi)^2\}]\) below,
is positive (when it vanishes --- a treated arm without residual
variance and without effect dispersion given \(S\) --- the ratio
diverges instead, which only strengthens the point). In setting (b), fix
the law of \((S, W)\) and the conditional laws of \(Y\) given
\((S, W, A)\), and let \(e_S = \lambda\,\tilde e_S\) with
\(\lambda \downarrow 0\), so that treatment becomes rare in every
stratum while \(\mu_a\), \(\sigma_a^2\), \(\tau\), \(V_0\), \(C_{10}\)
and \(\psi\) are unchanged. Then
\(p^2 V_{\mathrm{att}}(S) = \lambda\,E[\tilde e_S\{\sigma_1^2(S) + (\tau_S - \psi)^2\}] + O(\lambda^2)\)
and \(p^2\{V_{\mathrm{att}}(S \cup W) - V_{\mathrm{att}}(S)\}
= \lambda\,E[\tilde e_S\{V_0(S) - 2 C_{10}(S)\}] + O(\lambda^2)\), so
that \[
\lim_{\lambda \downarrow 0}
\frac{V_{\mathrm{att}}(S \cup W) - V_{\mathrm{att}}(S)}{V_{\mathrm{att}}(S)}
 \;=\; \frac{E[\tilde e_S\{V_0(S) - 2 C_{10}(S)\}]}
 {E[\tilde e_S\{\sigma_1^2(S) + (\tau_S - \psi)^2\}]}.
\] The control-arm term, with its weight \(e^2/(1-e)\), drops out at
leading order, and what survives of the increment is the
effect-dispersion loss with weight \(e\). In the family of Section 5 the
limit is \(v_w = \delta^2 q(1-q)\): scaling
\((e_0, e_1) = (0.168, 0.401)\) by \(0.1\), \(0.01\) and \(0.001\)
raises the penalty of Theorem 2 from 14.4 to 51.5, 55.8 and 56.2
percent, against the limit of 56.25 percent.

\textbf{Corollary 3 (when supplementation is safe).} In setting (b), if
\(\mu_1(S \cup W) - \mu_0(S \cup W)\) is \(\sigma(S)\)-measurable --- no
effect modification by \(W\) given \(S\) --- then
\(V_1(S) = V_0(S) = C_{10}(S)\) and (3) reduces to \[
V_{\mathrm{att}}(S \cup W) - V_{\mathrm{att}}(S)
 = -\frac{1}{p^2} E\Bigl[ \frac{e_S}{1-e_S}\, V_0(S) \Bigr] \le 0 :
\] absent modification, outcome-side supplementation always helps the
ATT, as it does the ATE.

\textbf{Corollary 4 (why the ATE is protected).} In setting (b), for the
ATE, \[
V_{\mathrm{ate}}(S \cup W) - V_{\mathrm{ate}}(S)
 = -E\Bigl[ \frac{1-e_S}{e_S} V_1(S) + \frac{e_S}{1-e_S} V_0(S)
  + 2 C_{10}(S) \Bigr] \le 0
\] always, because
\(\frac{1-e}{e} V_1 + \frac{e}{1-e} V_0 \ge 2\sqrt{V_1 V_0} \ge
2|C_{10}|\) pointwise (AM--GM and Cauchy--Schwarz). The ATT weights
\((e, e^2/(1-e))\) do not admit such a bound, and Corollary 2 shows the
failure is real.

\emph{Remark 4 (a plug-in diagnostic).} The sign functional in (3) is
estimable: fit \(\mu_0\) and \(\mu_1\) on \(S \cup W\) and on \(S\),
form the conditional variance and covariance of the fitted differences,
and average against \(e_S\). A negative estimate says the extension
helps; a positive one says it hurts. The diagnostic requires a
treated-arm regression \(\mu_1\) in addition to the \((e_S, \mu_0)\)
that a doubly robust ATT estimator fits, and it carries its own sampling
uncertainty --- with 185 treated units, as in Section 9, that
uncertainty is not small --- so it should be reported with an interval
and not used as an automatic selection rule. Section 9 does not estimate
it; it reports only a regression-difference sensitivity measure.

\section{No graphical criterion for the ATT}\label{sec-nongraphical}

For the ATE, the optimal graphically valid adjustment set is a function
of the graph alone: one set, \(O(G)\), is optimal within
\(\mathcal{A}(G)\) under \emph{every} distribution Markov to \(G\)
(\citeproc{ref-rotnitzkysmucler2020}{Rotnitzky and Smucler 2020};
\citeproc{ref-henckel2022}{Henckel, Perković, and Maathuis 2022}). We
now show that no such map can exist for the ATT.

Let \(G^{*}\) be the DAG with vertices \(\{Z, W, A, Y\}\) and edges
\(Z \to A\), \(Z \to Y\), \(W \to Y\), \(A \to Y\). Its graphically
valid adjustment sets are \(\{Z\}\) and \(\{Z, W\}\), and its
ATE-optimal set is
\(O(G^{*}) = \mathrm{pa}(Y) \setminus \{A\} = \{Z, W\}\). In the family
below, no other subset of \(\{Z, W\}\) is valid under any member
distribution, so \(\mathcal{A}(G^{*})\) and \(\mathcal{A}(P)\) coincide.

Consider the family \(\mathcal{F}\) of distributions: \(Z \sim
\mathrm{Bern}(r)\) and \(W \sim \mathrm{Bern}(q)\) independent;
\(A \mid Z, W \sim \mathrm{Bern}\{e(Z)\}\) with \(e(0) = e_0\),
\(e(1) = e_1\); and \[
Y \;=\; \gamma Z + \delta\,(1 - A)\,W + \theta A + \varepsilon,
\qquad \varepsilon \sim N(0, 1) \ \text{independent},
\] with \(r, q \in (0,1)\), \(e_0 \ne e_1 \in (0,1)\), and
\(\gamma, \delta \ne 0\). Every member is strictly positive and Markov
to \(G^{*}\); faithfulness to \(G^{*}\) holds outside a Lebesgue-null
set of parameter values, in particular at the two instances used below
(Appendix A.3).

\textbf{Lemma 3 (closed forms on \(\mathcal{F}\)).} Write
\(p = E\{e(Z)\}\), \(H = E\{\kappa(e(Z))\}\),
\(v_w = \delta^2 q (1-q)\). Then \[
V_{\mathrm{att}}(\{Z\}) = \frac{p + H\,(1 + v_w)}{p^{2}},
\qquad
V_{\mathrm{att}}(\{Z, W\}) = \frac{p + H + p\, v_w}{p^{2}},
\] so that \[
V_{\mathrm{att}}(\{Z,W\}) - V_{\mathrm{att}}(\{Z\})
 \;=\; \frac{v_w}{p^{2}}\,(p - H)
 \;=\; \frac{v_w}{p^{2}}\; E\Bigl[ \frac{e(Z)\{1 - 2e(Z)\}}{1 - e(Z)} \Bigr].
\tag{4}
\] For the ATE, \(V_{\mathrm{ate}}(\{Z,W\}) - V_{\mathrm{ate}}(\{Z\})
= -\,v_w\, E[e(Z)/\{1-e(Z)\}] < 0\) for every member of \(\mathcal{F}\).

\textbf{Theorem 2 (ATT-optimality is not graphical).} Let
\(P \in \mathcal{F}\) be faithful to \(G^{*}\) with \(e_0, e_1 < 1/2\)
and \(P' \in \mathcal{F}\) be faithful to \(G^{*}\) with
\(e_0, e_1 > 1/2\) (such \(P\) and \(P'\) exist; Appendix A.3). Then
\(P\) and \(P'\) are both Markov and faithful to \(G^{*}\), yet the
unique ATT-optimal valid adjustment set is \(\{Z\}\) under \(P\) and
\(\{Z, W\}\) under \(P'\). Consequently:

\begin{enumerate}
\def\labelenumi{(\roman{enumi})}
\item
  there is no map \((G, A, Y) \mapsto S(G)\) that outputs an ATT-optimal
  valid adjustment set for all distributions Markov (or Markov and
  faithful) to \(G\) --- in contrast to the ATE, for which \(O(G)\) is
  such a map;
\item
  the ATE-optimal set \(O(G^{*}) = \{Z, W\}\) is strictly ATT-suboptimal
  under \(P\), with exact penalty \((v_w / p^2)\, E[e(1-2e)/(1-e)] > 0\)
  given by (4).
\end{enumerate}

\emph{Proof.} Lemma 3 and the sign of \(e(1-2e)/(1-e)\) on \(e < 1/2\)
versus \(e > 1/2\); Appendix A.3 verifies validity, positivity, and the
faithfulness claims. The instance
\((e_0, e_1, r, q, \gamma, \delta, \theta) =
(0.168, 0.401, 0.5, 0.5, 1, 1.5, 1)\) gives
\(V_{\mathrm{att}}(\{Z\}) = 6.433\) versus
\(V_{\mathrm{att}}(\{Z,W\}) = 7.360\) --- a 14.4 percent penalty (6.96
percent in standard error) for using \(O(G^{*})\) --- while its
\(e \mapsto 1 - e\) mirror flips the ranking. \(\square\)

\textbf{Corollary 5 (what a graph can still provide).} The graph
determines the graphically valid class \(\mathcal{A}(G)\) and the
treatment-side ordering (Theorem 1(a)); the outcome-side ordering
additionally requires the sign functional in (3), which depends on the
law through \((e_S, V_0, C_{10})\). Any complete ATT selection procedure
therefore needs information beyond the graph, and a natural
implementation is \emph{two-stage} --- graphical pruning of instruments
and forbidden nodes, followed by a data- or assumption-driven resolution
of (3); Theorem 3 gives the no-modification condition under which the
second-stage comparison is immediate and \(O(G)\) itself is optimal
within the graphically valid class.

\section{When the ATE-optimal set is ATT-optimal}\label{sec-thm3}

Theorem 2 shows that no graph-only rule can be ATT-optimal in general. A
complementary question is under what additional condition the
ATE-optimal set retains its optimality. The answer is a sufficient
condition for the two estimands to coincide --- the absence of effect
modification by the covariates, which is not necessary for
\(\psi_{\mathrm{att}} = \psi_{\mathrm{ate}}\), since the two can agree
under effect modification that averages out.

We need three graphical facts, all from the ATE literature. Following
Rotnitzky and Smucler (\citeproc{ref-rotnitzkysmucler2020}{2020}) we
write \(Z\) for a generic valid adjustment set in this section, in
Section 10 and in Appendix A.4. Let \(G\) be a DAG over
\((Z_1, \dots, Z_J, A, Y)\) with \(Y\) a descendant of \(A\), let
\(\mathrm{cn}(A, Y, G)\) be the non-\(A\) nodes on causal paths from
\(A\) to \(Y\),
\(\mathrm{forb}(A, Y, G) = \mathrm{de}_G\{\mathrm{cn}(A,Y,G)\}
\cup \{A\}\), and
\(O(G) = \mathrm{pa}_G\{\mathrm{cn}(A, Y, G)\} \setminus
\mathrm{forb}(A, Y, G)\).

\textbf{Lemma 4} (\citeproc{ref-rotnitzkysmucler2020}{Rotnitzky and
Smucler 2020}, Lemmas 4--5 and Theorems 6--7;
\citeproc{ref-henckel2022}{Henckel, Perković, and Maathuis 2022},
Theorem 3 (Theorem 3.13 of the arXiv version) and Lemmas E.4--E.5 of its
supplement). For a point treatment, with validity meaning membership in
\(\mathcal{A}(G)\): (i) \(O(G)\) is a valid adjustment set; (ii) for
every valid \(Z\), \(A \perp_G \{O(G) \setminus Z\} \mid Z\) and
\(Y \perp_G \{Z \setminus O(G)\} \mid O(G), A\); (iii) if \(Z\) is valid
and \(A \perp_G B \mid Z\) then \(Z \cup B\) is valid, and if
\(Z \cup B\) is valid and \(Y \perp_G B \mid Z, A\) then \(Z\) is valid.

The restriction to \(\mathcal{A}(G)\) is not a formality. A set that is
valid at the law but not graphically valid need not satisfy (ii), and
Theorem 3 below fails for such competitors: with \(U, V\) independent
Bernoulli\((1/2)\), the single four-level covariate \(X = 2U + V\),
\(P(A = 1 \mid X) = 0.2\) when \(U = 0\) and \(0.8\) when \(U = 1\), and
\(Y(a) = V + a + \varepsilon\) with \(\varepsilon \sim N(0,1)\)
independent, the observed law is faithful to the DAG \(X \to A\),
\(X \to Y\), \(A \to Y\), the effect is constant, and \(O(G) = \{X\}\);
yet \(A \perp\!\!\!\perp (V, \varepsilon)\) makes the empty set valid at
this law, with \(V_{\mathrm{att}}(\varnothing) = 5\) against
\(V_{\mathrm{att}}(\{X\}) = 8.5\) by (1). The empty set is not in
\(\mathcal{A}(G)\), and its validity is not implied by the graph.

Say that a distribution exhibits \textbf{no effect modification} if
\(E\{Y(1) - Y(0) \mid Z_1, \dots, Z_J\} = E\{Y(1) - Y(0)\}\) almost
surely, that is, conditionally on the full covariate vector. Under this
condition \(\psi = \psi_{\mathrm{ate}}\), and for every valid \(S\),
\(\tau_S = E\{Y(1) - Y(0) \mid S\}\) is constant.

\textbf{Theorem 3 (ATT-optimality of \(O(G)\) under no effect
modification).} Let \(P\) be Markov to \(G\), satisfy the regularity
conditions of Section 2 for every \(Z \in \mathcal{A}(G)\), and exhibit
no effect modification. Then for every graphically valid adjustment set
\(Z \in \mathcal{A}(G)\), \[
\begin{aligned}
V_{\mathrm{att}}(Z) - V_{\mathrm{att}}(O)
 &= \frac{1}{p^{2}} E\Bigl[ \frac{e_Z}{1 - e_Z}\,
 \mathrm{var}\{\mu_0(Z \cup O) \mid Z\} \Bigr] \\
 &\quad + \frac{1}{p^{2}} E\Bigl[ \sigma_0^2(O)\,
 \bigl\{ E\bigl(\kappa(e_{Z \cup O}) \bigm| O\bigr) - \kappa(e_O) \bigr\} \Bigr]
 \;\ge\; 0,
\end{aligned}
\tag{5}
\] with \(O = O(G)\); hence \(O(G)\) is ATT-optimal within
\(\mathcal{A}(G)\). Equality holds if and only if
\(\mu_0(Z \cup O) = \mu_0(Z)\) almost surely, and \(e_{Z \cup O} = e_O\)
almost surely on \(\{\sigma_0^2(O) > 0\}\).

\emph{Proof.} Appendix A.4: supplement \(Z\) by \(O \setminus Z\)
(Theorem 1(b) with Corollary 3, licensed by Lemma 4(ii)--(iii)), then
delete \(Z \setminus O\) from the union (Theorem 1(a), licensed by Lemma
4(i)--(ii)). \(\square\)

Three remarks clarify this result. First, the proof follows exactly the
structure of the ATE optimality proof in Rotnitzky and Smucler
(\citeproc{ref-rotnitzkysmucler2020}{2020}) --- supplement, then delete,
through the union \(Z \cup O\) --- with the ATE increments replaced by
the ATT increments of Theorem 1; the no-modification condition supplies
the sign of the supplementation step, which is automatic for the ATE
(Corollary 4) but not for the ATT (Corollary 2). Second, the condition
cannot be dropped: every member of \(\mathcal{F}\) with \(\delta \ne 0\)
violates it, and under \(P \in \mathcal{F}\) with \(e < 1/2\) the set
\(O(G^{*})\) is strictly ATT-suboptimal. Third, no effect modification
makes the ATT and the ATE the same number but not the same estimation
problem: the two functionals have different bounds at the same set
(Table 1), and either may be the smaller one. When the condition is
maintained as a restriction on the model, the two functionals identify
the same parameter throughout the model and the analyst may use
whichever has the smaller bound; the efficient estimator under the
restriction, which would exploit it, is a different problem that we do
not treat. When the condition merely happens to hold at the law at hand,
an estimator of the ATE functional is not a regular estimator of the ATT
in the unrestricted model, and no such choice is licensed.

\section{Overlap-weighted and general weighted
estimands}\label{sec-wate}

The outcome-side calculus extends to the whole weighted class with a
single identity.

\textbf{Theorem 4 (outcome-side extension for a WATE).} Let \(S\) and
\(S \cup W\) be valid with \(W \perp\!\!\!\perp A \mid S\), and let
\(h\) be as in Lemma 2. Then \(\psi_h(S \cup W) = \psi_h(S)\),
\(D_{S \cup W} = D_S\), and, writing \(e = e_S\), \(h = h(e_S)\),
\(h' = h'(e_S)\), \[
\begin{aligned}
D_S^{2}\,\bigl\{ V_h(S \cup W) - V_h(S) \bigr\}
 &= E\Bigl[ V_1(S)\Bigl\{ h'^{2} e(1-e) - \frac{h^{2}(1-e)}{e} \Bigr\} \\
 &\qquad + V_0(S)\Bigl\{ h'^{2} e(1-e) - \frac{h^{2} e}{1-e} \Bigr\}
 - 2\, C_{10}(S)\,\bigl\{ h^{2} + h'^{2} e(1-e) \bigr\} \Bigr].
\end{aligned}
\tag{6}
\]

\emph{Proof.} Appendix A.5. \(\square\)

Setting \(h \equiv 1\) recovers Corollary 4 and \(h(e) = e\) recovers
(3) after multiplication by \(p^2\). The two coefficients in braces are
the \emph{arm-specific increments}: the first is the change in the bound
per unit of treated-arm-only predictive variance, and the second is the
change per unit of control-arm-only predictive variance. Their signs are
determined by the weight alone. For the ATE both are negative for every
\(e\). For the ATT the first vanishes identically and the second is
\(e(1-2e)/(1-e)\). For the overlap weights the two are mirror images.

\textbf{Corollary 6 (ATO thresholds).} For \(h(e) = e(1-e)\) the
arm-specific increments are \[
\text{treated-arm: } -\,e^{2}(1-e)(2 - 3e), \qquad
\text{control-arm: } \phantom{-}e(1-e)^{2}(1 - 3e).
\] Hence a covariate that predicts only the treated arm given \(S\) (so
\(V_0 = C_{10} = 0\)) \emph{harms} ATO efficiency if and only if
\(E[V_1(S)\, e_S^{2}(1-e_S)(2-3e_S)] < 0\): in particular whenever
\(e_S \ge 2/3\) almost surely and \(V_1(S) > 0\) on a subset of
\(\{e_S > 2/3\}\) of positive probability, and never when
\(e_S \le 2/3\) almost surely. A covariate that predicts only the
control arm harms it if and only if
\(E[V_0(S)\, e_S(1-e_S)^{2}(1-3e_S)] > 0\), with the mirror sufficient
condition \(e_S \le 1/3\) almost surely and \(V_0(S) > 0\) on a
positive-probability subset of \(\{e_S < 1/3\}\), and never when
\(e_S \ge 1/3\) almost surely. The aggregate sign is that of an
expectation, and mixed support can produce harm on either side of a
threshold: with two equally likely strata, \(e_S = (0.2, 0.8)\) and
\(V_0(S) = (10, 1)\), the second expectation is
\(\tfrac12(0.0512 \times 10 - 0.0448 \times 1) = 0.2336 > 0\). The
thresholds are exchanged under \(e \mapsto 1 - e\) together with the
arms, as the symmetry of the overlap weights requires; the ATT's
pointwise threshold of \(1/2\) for control-arm predictors and \(\infty\)
for treated-arm predictors is the asymmetric limit of the same
structure.

\begin{figure}

\centering{

\pandocbounded{\includegraphics[keepaspectratio]{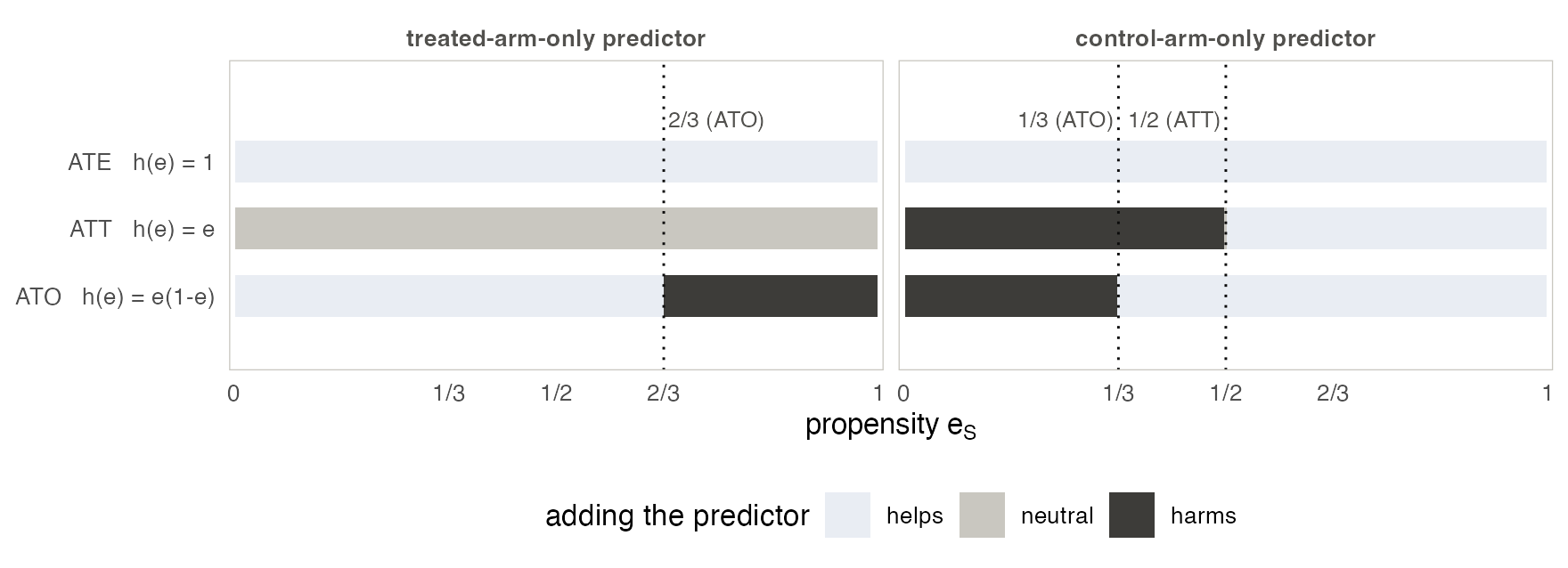}}

}

\caption{\label{fig-signmap}Pointwise sign of the outcome-side increment
for a covariate that predicts only one arm, by estimand and propensity
(Theorem 4, Corollary 6). For the ATE every extension helps; for the ATT
a treated-arm predictor is neutral and a control-arm predictor harms
below \(e = 1/2\); for the ATO the two thresholds are \(2/3\) and
\(1/3\). The aggregate sign at a law is that of the expectation of these
coefficients against the arm-specific predictive variances (Corollary
6).}

\end{figure}%

Corollary 6 concerns covariates that predict one arm only. The matrix
form of (6) settles the general case. Write \(K = h^2 + h'^2 e(1-e)\)
and let \(\Sigma(S)\) be the conditional covariance matrix of
\(\{\mu_1(S \cup W), \mu_0(S \cup W)\}\) given \(S\), with entries
\((V_1, C_{10}; C_{10}, V_0)\). Then the integrand of (6) is
\(\operatorname{tr}\{M(e_S)\,\Sigma(S)\}\) with \[
M(e) = \begin{pmatrix} K - h^{2}/e & -K \\ -K & K - h^{2}/(1-e) \end{pmatrix},
\qquad \det M(e) = -\,h(e)^{2}\,h'(e)^{2}.
\]

\textbf{Corollary 7 (constant weights are the only safe weights).} Let
\(h\) be continuously differentiable and positive on \([c, 1-c]\), and
let \(S\) be valid with \(e_S \in [c, 1-c]\). (a) If \(h' = 0\) on
\([c, 1-c]\), then \(M(e)\) is negative semidefinite for every \(e\),
and \(V_h(S \cup W) \le V_h(S)\) for every outcome-side extension
(\(W \perp\!\!\!\perp A \mid S\)) and every law of the outcome given
\((S, W, A)\). (b) If \(h'(e_S) \ne 0\) with positive probability, then
\(M(e_S)\) is indefinite there, and there exist an outcome-side
extension \(W\) and a law of the potential outcomes --- starting from
any law in which \(S\) is valid, with potential outcomes
\(Y^{\circ}(a)\), draw \(W\) conditionally on \(S\) as
Bernoulli\((1/2)\) independently of \((A, Y^{\circ}(0), Y^{\circ}(1))\),
and set \(Y(a) = Y^{\circ}(a) + v_a(S)(2W - 1)\); then \(S\) and
\(S \cup W\) are both valid, \(e_{S \cup W} = e_S\), and
\(\mu_a(S \cup W) = \mu_a(S) + v_a(S)(2W - 1)\) --- under which
\(V_h(S \cup W) > V_h(S)\). Hence, up to normalization, the ATE weights
are the only smooth positive weights for which outcome-side
supplementation never increases the bound, whatever the joint predictive
structure of the two arms.

\emph{Proof.} Appendix A.5. \(\square\)

Corollary 4 is the case (a); Corollary 2 and Corollary 6 exhibit the
covariance structures that realize (b) for the ATT and the ATO. The
corollary sharpens the message of Section 4: the protection the ATE
enjoys is not a matter of degree but of the derivative \(h'\), and every
propensity-dependent weight loses it.

Deletion is different in kind. For affine weights, adjusting for an
instrument changes the bound but not the estimand, and never for the
better: with \(h(e) = a + be\) the coefficient functions \(h^2/e\),
\(h^2/(1-e)\) and \(h^2 + h'^2 e(1-e)\) in (2) are convex, convex and
affine in \(e\), and \(D_S\) is unchanged, so the argument of Theorem
1(a) applies verbatim. For a nonaffine weight it can change both.

\textbf{Proposition 1 (estimand drift under instrument adjustment).} Let
\(S\) and \(S \cup W\) be valid with \(Y \perp\!\!\!\perp W \mid A, S\).
Then \(\tau_{S \cup W} = \tau_S\) and \[
\psi_h(S \cup W) - \psi_h(S)
 \;=\; \frac{E\bigl[ \{ E(h(e_{S \cup W}) \mid S) - h(e_S) \}\,\{\tau_S - \psi_h(S)\} \bigr]}
 {E\{h(e_{S \cup W})\}}.
\tag{7}
\] The drift vanishes for every law if and only if \(h\) is affine ---
in particular for the ATE, the ATT and the ATC --- and for a nonaffine
\(h\) it is nonzero for some law. At a given law, write
\(g(S) = E\{h(e_{S \cup W}) \mid S\} - h(e_S)\) for the Jensen gap; the
numerator of (7) is \(E[g\,\{\tau_S - \psi_h(S)\}]\), the expected
product of the gap and the effect centered at the \emph{\(h\)-weighted}
mean \(\psi_h(S)\), not the ordinary covariance of \(g\) and \(\tau_S\).
When \(h(e_S) > 0\) the distinction has a clean form: with \(Q_h\) the
law tilted by \(dQ_h = h(e_S)\,dP/E\{h(e_S)\}\), under which
\(E_{Q_h}(\tau_S) = \psi_h(S)\), and \(r = g/h(e_S)\) the relative gap,
\[
\psi_h(S \cup W) - \psi_h(S)
 = \frac{\mathrm{Cov}_{Q_h}(r, \tau_S)}{1 + E_{Q_h}(r)} .
\] The drift is thus zero exactly when the relative gap is uncorrelated
with the effect under the tilted law; the ordinary covariance is the
wrong diagnostic. With two equally likely strata, \(e_S = (0.2, 0.5)\),
\(\tau_S = (0, 1)\), and an independent binary instrument shifting each
propensity by \(\pm 0.1\), the overlap weights \(0.16\) and \(0.25\)
become \(0.15\) and \(0.24\) after averaging over the instrument, so the
absolute gap is \(-0.01\) in both strata and covaries with nothing, yet
the target moves from \(25/41\) to \(8/13\), a drift of
\(3/533 \approx 0.0056\); conversely, with \(e_S = (0.3, 0.7)\),
\(\tau_S = (0, 1)\) and the same instrument, the weights are \(0.21\) in
both strata and become \(0.20\) in both, the relative gaps are equal,
and \(\psi_h(S \cup W) = \psi_h(S) = 1/2\) although \(\tau_S\) varies.
For the concave overlap weight, \(g \le 0\) wherever \(W\) spreads \(e\)
within \(S\), so adjusting for an instrument lowers the unnormalized
weight of every stratum in which the instrument is informative; writing
\(\ell = -r \ge 0\) for the relative loss, the normalized share of a
stratum falls exactly when \(\ell\) exceeds its \(Q_h\)-average.

\emph{Proof.} Appendix A.6. \(\square\)

Proposition 1 has the following practical implication: for
overlap-weighted estimands, ``adjust for the outcome parents and drop
the instruments'' is not a variance recommendation but in general a
change of target, and two analysts who follow it with different valid
sets may estimate different quantities. An efficiency theory for the ATO
must therefore first fix a canonical weight --- the overlap weight
evaluated at a designated reference propensity, such as the full
pretreatment covariate propensity --- and compare adjustment sets for
the estimation of that fixed functional. We leave this to future work;
the functional so defined is a single parameter of the full law, and its
gradient and bound are not those of Lemma 2.

\section{Simulation study}\label{sec-sims}

The finite-sample study addresses two questions: whether the bound
orderings of Theorems 1--2 govern root mean squared error at moderate
\(n\), and whether influence-function standard errors are calibrated
when the adjustment set is deliberately chosen to be suboptimal for the
estimand.

\emph{Design.} We use the family \(\mathcal{F}\) of Theorem 2 with
\((r, q, \gamma, \delta, \theta) = (0.5, 0.5, 1, 1.5, 1)\), augmented by
a binary instrument \(I\) that shifts the treatment log-odds by
\(\pm 0.4\) within \(Z\). Design A is the rare-treatment member with
\((e_0, e_1) = (0.168, 0.401)\) at the midpoint of the instrument shift,
Design B the common-treatment member \((0.599, 0.832)\), its mirror
under \(e \mapsto 1 - e\) up to relabeling \(Z\); averaged over \(I\),
the within-\(Z\) propensities are \((0.175, 0.405)\) and
\((0.595, 0.825)\), which is why the exact \(\{Z,W\}\)-to-\(\{Z\}\)
bound ratio below is \(1.141\) rather than the \(1.144\) of Section 5.
We take \(n \in \{500, 2000\}\) and 1,000 replications. We use the
saturated nonparametric augmented inverse probability weighted (AIPW)
estimator for the ATT, with cell-wise propensities and control-arm
means. Its large-sample variance equals the bound in this discrete
design. The standard error is computed from the empirical variance of
the estimated influence function. A cell with no treated or no control
unit is dropped from that replication's estimator rather than the
replication being discarded; this occurred in 2 of the 1,000
replications of design A at \(n = 500\) for the set \(\{Z, W, I\}\) (a
cell of 62 and one of 59 control-only observations) and in no other cell
of the table. The point estimate is unaffected, since control-only cells
receive zero weight, and the standard error differs from the alternative
convention of retaining their zero influence-function contributions by
about 0.014 percent. The exact bounds \(V_{\mathrm{att}}(S)\) are
computed by enumerating the eight \((Z, W, I)\) cells
(\texttt{analysis/sim\_finite\_sample\_v2.R}), so the last two columns
of Table~\ref{tbl-sims} compare a Monte Carlo variance ratio, with its
standard error from a paired bootstrap over replications, with an exact
ratio. We compare four adjustment sets: \(\{Z\}\),
\(\{Z, W\} = O(G^*)\), \(\{Z, I\}\), and \(\{Z, W, I\}\).

\begin{longtable}[]{@{}
  >{\raggedright\arraybackslash}p{(\linewidth - 16\tabcolsep) * \real{0.0900}}
  >{\raggedright\arraybackslash}p{(\linewidth - 16\tabcolsep) * \real{0.0800}}
  >{\raggedright\arraybackslash}p{(\linewidth - 16\tabcolsep) * \real{0.1900}}
  >{\raggedright\arraybackslash}p{(\linewidth - 16\tabcolsep) * \real{0.0900}}
  >{\raggedright\arraybackslash}p{(\linewidth - 16\tabcolsep) * \real{0.0800}}
  >{\raggedright\arraybackslash}p{(\linewidth - 16\tabcolsep) * \real{0.0900}}
  >{\raggedright\arraybackslash}p{(\linewidth - 16\tabcolsep) * \real{0.1200}}
  >{\raggedright\arraybackslash}p{(\linewidth - 16\tabcolsep) * \real{0.1500}}
  >{\raggedright\arraybackslash}p{(\linewidth - 16\tabcolsep) * \real{0.1100}}@{}}
\caption{Finite-sample performance of nonparametric AIPW-ATT by
adjustment set (design A: rare treatment,
\((e_0, e_1) = (0.168, 0.401)\) at the instrument midpoint; design B:
its mirror), 1,000 replications
(\texttt{analysis/sim\_finite\_sample\_v2.R}). Monte Carlo standard
errors: bias 0.002--0.004, coverage 0.006--0.008, variance ratio in
parentheses (paired bootstrap over replications). The exact bound ratio
is \(V_{\mathrm{att}}(S)/V_{\mathrm{att}}(\{Z\})\) from the enumerated
design. At \(n = 500\), the Monte Carlo variances exceed the bounds by
6--11 percent under rare treatment; under common treatment, their
relative deviations from the bounds range from \(-1\) to \(+9\) percent.
At \(n = 2{,}000\), the relative deviations range from \(-6\) to \(+1\)
percent.}\label{tbl-sims}\tabularnewline
\toprule\noalign{}
\begin{minipage}[b]{\linewidth}\raggedright
Design
\end{minipage} & \begin{minipage}[b]{\linewidth}\raggedright
\(n\)
\end{minipage} & \begin{minipage}[b]{\linewidth}\raggedright
Set
\end{minipage} & \begin{minipage}[b]{\linewidth}\raggedright
Bias
\end{minipage} & \begin{minipage}[b]{\linewidth}\raggedright
SD
\end{minipage} & \begin{minipage}[b]{\linewidth}\raggedright
SE/SD
\end{minipage} & \begin{minipage}[b]{\linewidth}\raggedright
Coverage
\end{minipage} & \begin{minipage}[b]{\linewidth}\raggedright
Var. rel. to \(\{Z\}\) (MCSE)
\end{minipage} & \begin{minipage}[b]{\linewidth}\raggedright
Exact bound ratio
\end{minipage} \\
\midrule\noalign{}
\endfirsthead
\toprule\noalign{}
\begin{minipage}[b]{\linewidth}\raggedright
Design
\end{minipage} & \begin{minipage}[b]{\linewidth}\raggedright
\(n\)
\end{minipage} & \begin{minipage}[b]{\linewidth}\raggedright
Set
\end{minipage} & \begin{minipage}[b]{\linewidth}\raggedright
Bias
\end{minipage} & \begin{minipage}[b]{\linewidth}\raggedright
SD
\end{minipage} & \begin{minipage}[b]{\linewidth}\raggedright
SE/SD
\end{minipage} & \begin{minipage}[b]{\linewidth}\raggedright
Coverage
\end{minipage} & \begin{minipage}[b]{\linewidth}\raggedright
Var. rel. to \(\{Z\}\) (MCSE)
\end{minipage} & \begin{minipage}[b]{\linewidth}\raggedright
Exact bound ratio
\end{minipage} \\
\midrule\noalign{}
\endhead
\bottomrule\noalign{}
\endlastfoot
A & 500 & \(\{Z\}\) & 0.001 & 0.116 & 0.97 & 0.942 & 1.00 & 1.000 \\
A & 500 & \(\{Z,W\} = O\) & 0.002 & 0.126 & 0.96 & 0.944 & 1.18 (0.045)
& 1.141 \\
A & 500 & \(\{Z,I\}\) & 0.002 & 0.122 & 0.96 & 0.941 & 1.10 (0.019) &
1.072 \\
A & 500 & \(\{Z,W,I\}\) & 0.003 & 0.129 & 0.96 & 0.941 & 1.24 (0.049) &
1.187 \\
A & 2000 & \(\{Z\}\) & −0.003 & 0.055 & 1.02 & 0.959 & 1.00 & 1.000 \\
A & 2000 & \(\{Z,W\} = O\) & −0.004 & 0.060 & 1.00 & 0.958 & 1.20
(0.045) & 1.141 \\
A & 2000 & \(\{Z,I\}\) & −0.003 & 0.057 & 1.03 & 0.959 & 1.05 (0.018) &
1.072 \\
A & 2000 & \(\{Z,W,I\}\) & −0.004 & 0.061 & 1.01 & 0.960 & 1.22 (0.046)
& 1.187 \\
B & 500 & \(\{Z\}\) & −0.003 & 0.132 & 1.00 & 0.952 & 1.00 & 1.000 \\
B & 500 & \(\{Z,W\} = O\) & 0.000 & 0.119 & 0.99 & 0.946 & 0.82 (0.031)
& 0.788 \\
B & 500 & \(\{Z,I\}\) & −0.003 & 0.142 & 0.99 & 0.946 & 1.16 (0.031) &
1.140 \\
B & 500 & \(\{Z,W,I\}\) & −0.001 & 0.130 & 0.95 & 0.935 & 0.96 (0.045) &
0.878 \\
B & 2000 & \(\{Z\}\) & −0.002 & 0.065 & 1.02 & 0.954 & 1.00 & 1.000 \\
B & 2000 & \(\{Z,W\} = O\) & −0.001 & 0.058 & 1.02 & 0.947 & 0.79
(0.031) & 0.788 \\
B & 2000 & \(\{Z,I\}\) & −0.004 & 0.070 & 1.01 & 0.957 & 1.16 (0.024) &
1.140 \\
B & 2000 & \(\{Z,W,I\}\) & −0.002 & 0.062 & 1.01 & 0.951 & 0.89 (0.038)
& 0.878 \\
\end{longtable}

Three findings emerge. First, estimated biases were below 0.005 in every
cell, at most two Monte Carlo standard errors from zero, and within the
simulated designs the influence-function standard errors are calibrated
(SE/SD within 0.95--1.03, coverage 0.935--0.960 with a Monte Carlo
standard error of about 0.007): in these designs the adjustment-set
question is purely one of variance, as the theory frames it. Second, the
orderings are those of the bounds, and the magnitudes match them within
Monte Carlo error. Under rare treatment, adding the control-arm
predictor \(W\) --- that is, using the ATE-optimal set ---
\emph{inflates} the variance: the Monte Carlo ratios are 1.18 and 1.20
(standard error 0.045) against an exact bound ratio of 1.141; under
common treatment it reduces the variance, 0.82 and 0.79 (standard error
0.031) against 0.788. Third, the instrument raises the variance in every
cell (Theorem 1(a)): exact ratios 1.072 and 1.140, Monte Carlo
1.05--1.16. The joint addition of \(W\) and \(I\) is not the sum of the
two increments (exact ratios 1.187 and 0.878 against
\(1.141 + 1.072 - 1 = 1.213\) and \(0.788 + 1.140 - 1 = 0.928\)); the
increments of Theorem 1 are conditional on the set they extend.

A companion random-structure scan (Appendix B) generates four binary
covariates with randomly assigned parent sets and arm-asymmetric effect
modification, and compares \(\arg\min_S V_{\mathrm{att}}(S)\) with
\(\arg\min_S V_{\mathrm{ate}}(S)\) over the graphically valid sets, the
bounds computed exactly by enumerating the sixteen covariate cells. Of
60 draws, 49 had a treatment parent, an outcome parent and at least two
valid sets; in 10 of them the two minimizers differed and the exact ATT
penalty for using the ATE-optimal set exceeded 1 percent, the threshold
at which the scan counts a divergence. At this seed every structure with
distinct minimizers cleared the threshold, so 10 is also the number of
structures with any positive penalty; the penalties range from 2.9 to
25.9 percent (median 14.7 percent). Treatment prevalence in the 49
structures ranges from 0.19 to 0.56. The frequency is a property of the
generating distribution stated in Appendix B and is offered as evidence
that the phenomenon of Theorem 2 is not confined to a knife-edge, not as
a rate for any population of applications; the two explicit laws of
Theorem 2, not the frequency, establish the result.

\section{Empirical illustration: the LaLonde ATT}\label{sec-application}

The National Supported Work data (\citeproc{ref-lalonde1986}{LaLonde
1986}) provide a canonical setting for illustrating ATT estimation. We
use the experimental subsample constructed by Dehejia and Wahba
(\citeproc{ref-dehejiawahba1999}{1999}) --- 185 trainees and 260
experimental controls, for whom 1974 earnings are available --- together
with the 15,992 Current Population Survey controls (LaLonde's CPS-1
comparison group); the composite sample is that of Dehejia and Wahba
(\citeproc{ref-dehejiawahba1999}{1999}; see also
\citeproc{ref-dehejiawahba2002}{Dehejia and Wahba 2002}), and all
earnings are in 1982 dollars. In the experimental subsample the
difference in mean 1978 earnings between trainees and controls is
\$1,794, the benchmark of Dehejia and Wahba
(\citeproc{ref-dehejiawahba1999}{1999}); its unpooled
difference-in-means standard error, recomputed from the subsample, is
\$671 (the published figure of \$633 uses the pooled convention). The
treated participants constitute 1.1 percent of the composite sample, but
this mixing proportion is not the conditional propensity that enters the
bound: under the full-covariate logit fitted below, the propensities of
the 185 treated units have median 0.34, mean 0.32 and interquartile
range 0.11--0.57, so the ATT bound is integrated over strata in which
\(e_S\) lies mostly between 0.1 and 0.6, not near 0.01.

The illustration assesses the sensitivity of the ATT estimate and its
standard error to the deletion of single covariates; it does not
estimate the sign functional of (3), and it is not a test of the theory.
We estimate the ATT by AIPW with a logistic propensity model and a
linear control-arm outcome model, both on the same adjustment set, with
propensities truncated at 0.9 for controls (the truncation does not
bind: under the full model the largest control propensity is 0.64) and
influence-function standard errors. The full set is the ten covariates
of the Dehejia--Wahba file (age, education, Black, Hispanic, married, no
degree, 1974 and 1975 earnings, and their zero indicators), entered
linearly --- not the richer propensity specifications of Dehejia and
Wahba (\citeproc{ref-dehejiawahba1999}{1999},
\citeproc{ref-dehejiawahba2002}{2002}); we then drop each covariate in
turn and report the estimate, its standard error, and a
control-prediction sensitivity measure: the mean over the treated units
\(i\) of
\(\{\hat\mu_{0,\text{full}}(Z_i) - \hat\mu_{0,\text{reduced}}(Z_i)\}^2\),
with \(Z_i\) the covariate vector of unit \(i\), as a percentage of the
treated residual variance
\(\widehat{\mathrm{var}}(Y - \hat\mu_{0,\text{full}} \mid A = 1) = 67.7\)
(in units of \$1,000\(^2\)). This measure summarizes the squared changes
in fitted control-outcome predictions for the treated units when a
covariate is dropped; it is not the conditional variance \(V_0\) of (3),
which would be weighted by the propensity and accompanied by \(C_{10}\).
Validity of the reduced sets is not established by any of these numbers,
and the parametric nuisances place the estimator outside the class of
Section 3.1; the table describes what happens to the estimate and its
standard error under each deletion, not which set is valid
(\texttt{analysis/apply\_lalonde\_att\_v2.R}).

\Needspace*{15\baselineskip}

\begin{longtable}[]{@{}
  >{\raggedright\arraybackslash}p{(\linewidth - 10\tabcolsep) * \real{0.3000}}
  >{\raggedright\arraybackslash}p{(\linewidth - 10\tabcolsep) * \real{0.1100}}
  >{\raggedright\arraybackslash}p{(\linewidth - 10\tabcolsep) * \real{0.0800}}
  >{\raggedright\arraybackslash}p{(\linewidth - 10\tabcolsep) * \real{0.1000}}
  >{\raggedright\arraybackslash}p{(\linewidth - 10\tabcolsep) * \real{0.1400}}
  >{\raggedright\arraybackslash}p{(\linewidth - 10\tabcolsep) * \real{0.2700}}@{}}
\caption{The LaLonde ATT by adjustment set (AIPW, ten covariates entered
linearly, 185 NSW trainees and 15,992 CPS-1 controls). Dehejia--Wahba
experimental subsample benchmark \$1,794, unpooled difference-in-means
standard error \$671 recomputed from that subsample (published pooled
standard error \$633); 1982 dollars.}\label{tbl-lalonde}\tabularnewline
\toprule\noalign{}
\begin{minipage}[b]{\linewidth}\raggedright
Adjustment set
\end{minipage} & \begin{minipage}[b]{\linewidth}\raggedright
ATT (\$1,000)
\end{minipage} & \begin{minipage}[b]{\linewidth}\raggedright
SE
\end{minipage} & \begin{minipage}[b]{\linewidth}\raggedright
ΔSE (\%)
\end{minipage} & \begin{minipage}[b]{\linewidth}\raggedright
Δ estimate (\$1,000)
\end{minipage} & \begin{minipage}[b]{\linewidth}\raggedright
Control-prediction sensitivity (\% of treated residual variance)
\end{minipage} \\
\midrule\noalign{}
\endfirsthead
\toprule\noalign{}
\begin{minipage}[b]{\linewidth}\raggedright
Adjustment set
\end{minipage} & \begin{minipage}[b]{\linewidth}\raggedright
ATT (\$1,000)
\end{minipage} & \begin{minipage}[b]{\linewidth}\raggedright
SE
\end{minipage} & \begin{minipage}[b]{\linewidth}\raggedright
ΔSE (\%)
\end{minipage} & \begin{minipage}[b]{\linewidth}\raggedright
Δ estimate (\$1,000)
\end{minipage} & \begin{minipage}[b]{\linewidth}\raggedright
Control-prediction sensitivity (\% of treated residual variance)
\end{minipage} \\
\midrule\noalign{}
\endhead
\bottomrule\noalign{}
\endlastfoot
Full (10 covariates) & 1.495 & 0.678 & --- & --- & --- \\
drop age & 1.563 & 0.667 & −1.6 & +0.068 & 0.845 \\
drop education & 1.510 & 0.677 & −0.1 & +0.015 & 0.123 \\
drop married & 1.576 & 0.673 & −0.7 & +0.081 & 0.007 \\
drop no degree & 1.511 & 0.671 & −1.0 & +0.016 & 0.028 \\
drop Hispanic & 1.499 & 0.677 & −0.1 & +0.004 & 0.004 \\
drop 1974 earnings & 1.532 & 0.675 & −0.4 & +0.037 & 1.625 \\
drop zero-1975 indicator & 1.467 & 0.671 & −1.0 & −0.028 & 0.578 \\
drop zero-1974 indicator & 1.310 & 0.663 & −2.2 & −0.185 & 0.009 \\
drop 1975 earnings & 0.962 & 0.709 & +4.6 & −0.533 & 4.006 \\
drop Black & 0.314 & 0.618 & −8.8 & −1.181 & 0.620 \\
\end{longtable}

Two descriptive observations follow. First, the full-set estimate,
\$1,495 (standard error \$678), is within half a standard error of the
experimental benchmark, which is itself an estimate with a standard
error of about \$670. Deleting 1975 earnings or the Black indicator
changes the estimate by \$530 and \$1,180, respectively; deleting the
zero-1974 indicator changes it by \$185; these deletions materially
change the estimate and would warrant substantive investigation before
any of them were adopted. We retain the full specification as the
reference, noting that neither agreement with the benchmark nor a
smaller standard error establishes the validity of a reduced set. The
other seven deletions move the estimate by at most \$81. Second, the
seven leave-one-out specifications that leave the estimate essentially
unchanged have estimated standard errors between \(0.1\) and \(1.6\)
percent below the full model's; the three deletions that move the
estimate also change the fitted propensity model materially (the dropped
covariate has an absolute Wald \(z\)-statistic of 5 to 16 in the full
propensity logit) and change the standard error by \(-2.2\), \(+4.6\)
and \(-8.8\) percent. These are specification-sensitivity results, not
estimates of the increment (3): the conditional independences
\(W \perp\!\!\!\perp A \mid S\) that Theorem 1(b) requires are not
established for any of the ten covariates, the last column of
Table~\ref{tbl-lalonde} is a regression-difference measure --- a
treated-sample mean squared difference between two fitted control
regressions, without the propensity weights of (3), without \(C_{10}\),
and without the denominator of the bound --- and a change in the
standard error after deleting a covariate mixes the efficiency mechanism
with the change in the fitted weights and in the validity of the set.
With the other nine covariates retained, the mean squared change in
fitted control-outcome predictions for the treated units amounts to at
most 4.0 percent of the treated residual variance of earnings (when 1975
earnings is dropped), and to less than one percent for most deletions.
Earnings are noisy relative to the predictive contribution of any single
covariate (residual standard deviation about \$8,200). The small mixing
proportion alone does not imply that an ATT efficiency increment is
negligible --- (3) weights the control-arm gain by \(e^2/(1-e)\) but the
effect-dispersion loss by \(e\), so a small propensity does not shrink
the increment relative to the bound (Remark 3) --- and the fitted
treated-unit propensities are centered near 0.3; this illustration does
not identify the mechanism behind the observed standard-error changes.
In this example substantive validity arguments should take precedence in
choosing the set, and efficiency arguments imported from the ATE ---
adjust for every outcome predictor --- have little to add. An empirical
assessment of the mechanism would involve estimating the components of
(3), together with their uncertainty, for a few prespecified extensions
whose conditional-independence assumptions are defensible. We do not
attempt that assessment here.

\section{Hidden variables}\label{sec-hidden}

Suppose now that the causal DAG contains unobserved nodes and that
adjustment sets are drawn from the observed pretreatment covariates.
Theorems 1, 2 and 4 and Proposition 1 are statements about the observed
law of \((S, A, Y)\) under Definition 1 and about conditional
independencies among observed variables; they refer to the graph only
through validity and hold verbatim in the hidden-variable setting. The
graphical content is confined to Theorem 3, whose proof uses exactly
three properties of \(O(G)\) (Lemma 4). Abstracting them yields the
hidden-variable version.

\textbf{Theorem 5 (hidden variables).} Let \(\mathcal{Z}\) be a class of
observed adjustment sets, each valid at \(P\), let \(P\) satisfy the
regularity conditions of Section 2 for every member of \(\mathcal{Z}\)
and exhibit no effect modification, and suppose there is
\(O^{*} \in \mathcal{Z}\) such that for every \(Z \in \mathcal{Z}\): (i)
\(O^{*} \setminus Z \perp\!\!\!\perp A \mid Z\) and \(Z \cup O^{*}\) is
valid; (ii) \(Y \perp\!\!\!\perp Z \setminus O^{*} \mid A, O^{*}\). Then
\(V_{\mathrm{att}}(Z) - V_{\mathrm{att}}(O^{*})\) equals the right-hand
side of (5) with \(O^{*}\) in place of \(O\), and \(O^{*}\) is
ATT-optimal within \(\mathcal{Z}\).

\emph{Proof.} The proof of Theorem 3 (Appendix A.4) uses only (i), (ii),
and the validity of \(O^{*}\). \(\square\)

The premises are graphical facts about a class of sets, and the ATE
literature supplies them for specific classes. Smucler, Sapienza, and
Rotnitzky (\citeproc{ref-smuclersapienza2022}{2022}) show that when
every observed variable is an ancestor of \(A\) or \(Y\), or when there
are no hidden variables, their set \(O\) satisfies the separations
\(Y \perp_G Z \setminus O \mid A, O\) and
\(A \perp_G O \setminus Z \mid Z\) for every other graphically valid
observed set \(Z\), and that analogous sets satisfy them within the
classes of minimal and of minimum-cardinality graphically valid sets in
general; under the causal Markov assumption these separations deliver
(i)--(ii) for the corresponding class \(\mathcal{Z}\), and Theorem 5
then transfers the ATE optimality of those sets to the ATT under no
effect modification. Runge (\citeproc{ref-runge2021}{2021})
characterizes, by necessary and sufficient graphical conditions, when a
graphically optimal set exists in an \emph{information-theoretic} sense,
an ordering that coincides with the variance ordering for estimators
satisfying his Assumption 2 (the ordinary least squares estimator in
Gaussian linear models being the verified case); that characterization
does not by itself supply (i)--(ii) for the nonparametric bound, and his
Example D exhibits a graph in which the information-optimal set and a
competing valid set satisfy neither separation. When no set with
(i)--(ii) exists for the class of interest, Theorem 5 is silent, while
Theorems 1, 2 and 4 continue to order any two valid sets they compare.
The two nonexistence phenomena --- the ATE's, from hidden variables, and
the ATT's, from effect modification (Theorem 2) --- are logically
independent and can co-occur.

\section{Discussion}\label{sec-discussion}

\emph{For practice.} Three rules follow from the results, all of them
statements about valid extensions of a valid set, so that the choice
they inform is between sets that both identify the ATT. Among valid
extensions satisfying \(W \perp\!\!\!\perp A \mid S\), when the
propensity is below one half in the strata represented in the treated
population, do not simply apply the ATE adjustment rule: a purely
treated-arm predictor is efficiency-neutral for the unrestricted ATT
functional (Corollary 1), a purely control-arm predictor increases the
bound under the sign conditions of Corollary 2, and the covariates worth
adding are those that predict \emph{both} arms in the same direction
(large positive \(C_{10}\)); the plug-in diagnostic of Remark 4 can help
distinguish these covariate roles using the observed data, subject to
the sampling uncertainty noted there. A covariate that satisfies the
outcome exclusion \(Y \perp\!\!\!\perp W \mid A, S\) of Theorem 1(a) and
leaves the set valid --- an instrument, in the graphical case --- should
be excluded for the ATT, and, by the same Jensen argument, for every
affine weight; for nonaffine weights such as the overlap weights its
exclusion can change the target (Proposition 1), so the target must be
fixed before the exclusion is described as an efficiency gain. When the
assumption of no effect modification is credible, the ATE-optimal set
may be used for the ATT with the guarantee of Theorem 3 among
graphically valid sets --- a case in which the ATT and ATE functionals
identify the same number with different bounds; when the condition is
maintained as a model restriction the analyst may use whichever has the
smaller bound, although the efficient estimator under the restriction is
a different problem that we do not treat.

\emph{For theory.} The results place the graphical optimal-adjustment
program in relation to the older efficiency literature on the ATT. It
was already known that adjusting for an additional covariate can reduce
precision when estimating the ATT: White and Lu
(\citeproc{ref-whitelu2011}{2011, secs. II.A--D}) compare the
set-indexed ATT bounds of Section 3 for a valid set and a valid superset
under propensity exclusion and note that the difference can take either
sign, Kitagawa and Muris (\citeproc{ref-kitagawamuris2016}{2016}) find
the same nonmonotonicity in parametric propensity submodels and note
that homogeneous effects remove it, and Hahn
(\citeproc{ref-hahn2004}{2004}) shows what a maintained propensity
restriction contributes to efficiency. We re-express that comparison in
conditional moments and use it to answer a question these works do not
ask, whether an adjustment set determined by the graph can be optimal
uniformly over the compatible laws. We obtain both a negative and a
positive result. No graph-only criterion can exist in general because
two faithful laws on one DAG can have different ATT-optimal sets.
However, \(O(G)\) remains optimal among graphically valid sets under no
effect modification. The ATE's graph-only optimality is not a generic
property of causal estimands but a consequence of its constant weights,
which admit an arithmetic--geometric-mean bound (Corollary 4) that no
smooth propensity-dependent weight admits (Corollary 7). For affine
weights the graph supplies half of the answer --- the graphically valid
class and the treatment-side ordering --- and the other half is a sign
functional of the law; for nonaffine weights, treatment-side comparisons
must first address the target drift of Proposition 1. The WCDE theory of
Lin, Guo, and Gan (\citeproc{ref-linguogan2025}{2025}), whose
mediator-marginal weights do not depend on the propensity, obtains a
graph-only optimal set in its own setting; we conjecture, but have not
shown, that the distinction between graph-only and law-dependent
optimality is governed by the derivative \(h'\) of the weight with
respect to the propensity score.

\emph{Limitations and extensions.} The bounds are nonparametric and
asymptotic; in finite samples with parametric nuisances the ordering can
be perturbed by model misspecification, which is a separate question
from the one asked here. Hidden variables are handled in Section 10 by
abstracting the three graphical properties Theorem 3 uses; the existence
question they raise is inherited from the ATE theory. Time-dependent
treatments, where the supplement-then-delete architecture does not by
itself yield uniform optima for the ATE
(\citeproc{ref-rotnitzkysmucler2020}{Rotnitzky and Smucler 2020, sec.
5.2}) and further graphical conditional independencies must be exploited
(\citeproc{ref-adenyo2025}{Adenyo et al. 2025}), are the natural next
case. Finally, the canonical-weight efficiency theory for the ATO
sketched after Proposition 1 is, in our view, the most useful open
problem the paper leaves.

\section*{Appendix A: Proofs}\label{sec-proofs}
\addcontentsline{toc}{section}{Appendix A: Proofs}

\subsection*{A.1 Proof of Lemma 1}\label{a.1-proof-of-lemma-1}
\addcontentsline{toc}{subsection}{A.1 Proof of Lemma 1}

Consider the nonparametric model for the observation \((S, A, Y)\) with
density \(f(s)\, \eta(s)^{a} \{1-\eta(s)\}^{1-a} f_a(y \mid s)\),
\(\eta = e_S\). Scores decompose as
\(g(s, a, y) = g_S(s) + g_{A}(a, s) + g_{Y}(y \mid a, s)\) with the
usual mean-zero restrictions. Write \(\chi = N/D\) with
\(N = E\{\eta\, \tau_S\}\) and \(D = p = E\{\eta\}\). Differentiating
along a submodel with score \(g\) gives
\(\dot\eta(s) = E[(A - \eta) g_A \mid s]\),
\(\dot\tau(s) = E[(Y - \mu_1) g_Y \mid A=1, s] - E[(Y - \mu_0) g_Y \mid A=0, s]\),
and \[
\dot\chi = \frac{1}{p}\Bigl( E[\eta(\tau_S - \psi) g_S]
 + E[(\tau_S - \psi)\dot\eta] + E[\eta\, \dot\tau] \Bigr).
\] For a candidate influence function of the form
\(\varphi = p^{-1}(\varphi_1 + \varphi_2 + \varphi_3)\), we match the
components as follows: \(E[\varphi_1 g_S]\) requires
\(\varphi_1 = \eta(s)(\tau_S - \psi)\);
\(E[(\tau_S - \psi)\dot\eta] = E[(\tau_S - \psi)(A - \eta) g_A]\) gives
\(\varphi_2 = (\tau_S - \psi)(A - \eta)\); and
\(E[\eta \dot\tau] = E[\{A (Y - \mu_1) - \tfrac{\eta}{1-\eta}(1-A)(Y-\mu_0)\} g_Y]\)
gives
\(\varphi_3 = A(Y - \mu_1) - \tfrac{\eta}{1-\eta}(1-A)(Y - \mu_0)\).
Summing, \(\varphi = p^{-1}[ A\{Y - \mu_0 - \psi\}
- \tfrac{\eta}{1-\eta}(1-A)\{Y - \mu_0\} ]\). The matching is exact
because the cross pairings vanish: each \(\varphi_k\) has zero
conditional mean with respect to the conditioning variable of the other
two score components (e.g. \(E[\varphi_3 \mid A, S] = 0\) eliminates its
pairings with \(g_S\) and \(g_A\), and \(\varphi_1, \varphi_2\) are free
of \(Y\) while \(E[g_Y \mid A, S] = 0\)). Since \(\varphi\) itself lies
in the tangent space, it is the \emph{efficient} influence function. To
compute its second moment, note that on \(\{A=1\}\),
\(E[A(Y - \mu_0 - \psi)^2]
= E[\eta\{\sigma_1^2 + (\tau_S - \psi)^2\}]\); on \(\{A=0\}\),
\(E[\tfrac{\eta^2}{(1-\eta)^2}(1-A)(Y-\mu_0)^2]
= E[\tfrac{\eta^2}{1-\eta}\sigma_0^2]\); the cross term vanishes since
\(A(1-A) = 0\). Equation (1) follows. \(\square\)

\subsection*{A.2 Proof of Theorem 1}\label{a.2-proof-of-theorem-1}
\addcontentsline{toc}{subsection}{A.2 Proof of Theorem 1}

Both parts compare (1) at \(S \cup W\) and at \(S\); note \(p\) and
\(\psi\) are functionals of the estimand, not of the set, so only the
three integrand components change.

\textbf{(a).} The hypothesis \(Y \perp W \mid A, S\) says that the
conditional law of \(Y\) given \((A, S, W)\) does not depend on \(W\);
hence \(\mu_a(S \cup W) = \mu_a(S)\) and
\(\sigma_a^2(S \cup W) = \sigma_a^2(S)\) for \(a = 0, 1\), and
\(\tau_{S \cup W} = \tau_S\). (Both sets being valid, these observed-arm
quantities are the potential-outcome quantities under either set, so the
two functionals identify the same \(\psi\).) In
\(p^2 V_{\mathrm{att}}(S \cup W)\), condition the first and third terms
on \(S\):
\(E[e_{S\cup W} \sigma_1^2(S)] = E[\sigma_1^2(S)\, E(e_{S\cup W} \mid S)]
= E[e_S \sigma_1^2(S)]\) by the tower property, and likewise for
\(e_{S \cup W}(\tau_S - \psi)^2\). The middle terms differ by
\(E[\sigma_0^2(S)\{E(\kappa(e_{S\cup W}) \mid S) - \kappa(e_S)\}]\), and
\(E(e_{S \cup W} \mid S) = e_S\) with \(\kappa\) strictly convex gives
the inequality and its equality condition by conditional Jensen.
\(\square\)

\textbf{(b).} \(W \perp A \mid S\) gives \(e_{S \cup W} = e_S\);
moreover the conditional law of \(W\) given \((S, A = a)\) equals that
given \(S\), so for \(a = 0, 1\), \[
\sigma_a^2(S) = E\{\sigma_a^2(S \cup W) \mid S\} + V_a(S), \qquad
\tau_S = E(\tau_{S \cup W} \mid S),
\] by the conditional variance decomposition, and
\(E\{(\tau_{S\cup W} - \psi)^2 \mid S\} = (\tau_S - \psi)^2 +
\mathrm{var}(\tau_{S \cup W} \mid S)\) with
\(\mathrm{var}(\tau_{S\cup W} \mid S) = V_1 + V_0 - 2 C_{10}\).
Substituting into
\(p^2\{V_{\mathrm{att}}(S \cup W) - V_{\mathrm{att}}(S)\}\) --- the
weights \(e_S\) and \(\kappa(e_S)\) are \(\sigma(S)\)-measurable and
pass through the conditional expectations --- gives \[
p^2 \Delta
= E\bigl[ -e_S V_1 - \kappa(e_S) V_0 + e_S (V_1 + V_0 - 2C_{10}) \bigr]
= E\bigl[ \{e_S - \kappa(e_S)\} V_0 - 2 e_S C_{10} \bigr],
\] and \(e - \kappa(e) = e(1-2e)/(1-e)\) gives (3). Equivalently, under
\(W \perp A \mid S\) the two influence functions differ by
\(\varphi_S - \varphi_{S \cup W} = (A - e_S)\, m / \{p(1 - e_S)\}\) with
\(m = \mu_0(S \cup W) - \mu_0(S)\), the \(\mu_1\) terms canceling, and
\(V_{\mathrm{att}}(S) - V_{\mathrm{att}}(S \cup W)
= E[(\varphi_S - \varphi_{S\cup W})^2] + 2E[(\varphi_S - \varphi_{S\cup W})\varphi_{S \cup W}]
= p^{-2} E[e_S V_0/(1 - e_S)] + 2 p^{-2} E[e_S (C_{10} - V_0)]\), which
is (3) with the sign reversed. \(\square\)

\emph{Corollaries.} 1: \(\mu_0(S \cup W) = \mu_0(S)\) makes
\(V_0 = C_{10} = 0\) in (3); the ATE statement follows from the display
in Corollary 4 with \(V_0 = C_{10} = 0\). 2: sign inspection of (3). 3:
\(\mu_1 - \mu_0\) \(\sigma(S)\)-measurable forces
\(\mu_1(S\cup W) = \mu_0(S \cup W) + g(S)\) with \(g(S) = \tau_S\),
hence \(V_1 = V_0 = C_{10}\), and the integrand of (3) becomes
\(e V_0 \{(1-2e)/(1-e) - 2\} = -e V_0/(1-e)\), yielding the display. 4:
the ATE increment follows by the same conditional decompositions applied
to \(V_{\mathrm{ate}}\), and the bound from
\(\tfrac{1-e}{e}V_1 + \tfrac{e}{1-e}V_0 \ge 2\sqrt{V_1 V_0}\) (AM--GM)
and \(|C_{10}| \le \sqrt{V_1 V_0}\) (Cauchy--Schwarz). \(\square\)

\subsection*{A.3 Proof of Lemma 3 and Theorem
2}\label{a.3-proof-of-lemma-3-and-theorem-2}
\addcontentsline{toc}{subsection}{A.3 Proof of Lemma 3 and Theorem 2}

\emph{Validity.} In \(G^{*}\) the only back-door path from \(A\) to
\(Y\) is \(A \leftarrow Z \to Y\); hence a subset of \(\{Z, W\}\) is
graphically valid iff it contains \(Z\):
\(\mathcal{A}(G^{*}) = \{\{Z\}, \{Z, W\}\}\). The two sets omitting
\(Z\) are also invalid at every member of \(\mathcal{F}\), so that
\(\mathcal{A}(P) = \mathcal{A}(G^{*})\): with
\(Y(0) = \gamma Z + \delta W + \varepsilon\) and \(W \perp (Z, A)\), \[
\begin{aligned}
E\{Y(0) \mid A = 1, W\} - E\{Y(0) \mid A = 0, W\}
&= \gamma\{E(Z \mid A = 1) - E(Z \mid A = 0)\} \\
&= \frac{\gamma\, r(1-r)(e_1 - e_0)}{p(1-p)},
\end{aligned}
\] which is nonzero because \(\gamma \ne 0\), \(e_0 \ne e_1\) and
\(r \in (0, 1)\); hence \(Y(0) \not\perp A \mid W\), and the same
display without \(W\) shows \(Y(0) \not\perp A\). Overlap holds for the
valid sets with \(c = \min(e_0, e_1, 1-e_0, 1-e_1)\).

\emph{Markov property and faithfulness.} Every member factorizes as
\(P(z)\,P(w)\,P(a \mid z)\,f(y \mid z, w, a)\) and is therefore Markov
to \(G^{*}\). Faithfulness to \(G^{*}\) is the conjunction of twenty
conditional-dependence statements: of the 24 statements
\(X \perp V \mid C\) with \(\{X, V\} \subset \{Z, W, A, Y\}\) and \(C\)
a subset of the remaining two variables, \(G^{*}\) \(d\)-separates
exactly four --- \(W \perp Z\), \(W \perp Z \mid A\), \(W \perp A\),
\(W \perp A \mid Z\) --- and these hold in every member because
\(W \perp (Z, A)\) by construction; the other twenty are \(d\)-connected
and must fail. Fifteen of them fail at every parameter point, by
inspection of a conditional mean, variance or log odds ratio:
\(Z\)--\(A\) and \(Z\)--\(A \mid W\), since \(e_0 \ne e_1\);
\(Z\)--\(Y \mid A\) and \(Z\)--\(Y \mid A, W\), since
\(E(Y \mid Z, A, W)\) has slope \(\gamma \ne 0\) in \(Z\);
\(Z\)--\(Y \mid W\), since the slope of \(E(Y \mid Z, W = w)\) in \(Z\)
is \(\gamma + (\theta - \delta w)(e_1 - e_0)\) for \(w = 0, 1\), and the
two cannot both vanish because \(\delta(e_1 - e_0) \ne 0\);
\(W\)--\(Y\), \(W\)--\(Y \mid Z\), \(W\)--\(Y \mid A\) and
\(W\)--\(Y \mid A, Z\), since the slope of \(E(Y \mid W, \cdot)\) in
\(W\) is \(\delta(1-p)\), \(\delta\{1 - e(Z)\}\) or \(\delta(1-A)\),
nonzero (on \(A = 0\) in the last case); \(A\)--\(Y \mid Z\), since
\(\mathrm{var}(Y \mid A=1, Z) = 1 \ne 1 + v_w =
\mathrm{var}(Y \mid A=0, Z)\); \(A\)--\(Y \mid Z, W\) and
\(A\)--\(Y \mid W\), since
\(E(Y \mid A=1, \cdot) - E(Y \mid A=0, \cdot)\) equals
\(\theta - \delta W\) and \(\gamma\{E(Z \mid A=1) - E(Z \mid A=0)\} +
\theta - \delta W\) respectively, neither of which vanishes for both
values of \(W\); \(Z\)--\(W \mid A, Y\), since given \(A = 0\) the log
odds ratio of \((Z, W)\) at any \(y\) equals \(-\gamma\delta \ne 0\);
\(W\)--\(A \mid Z, Y\), since given \(Z = z\) the log odds ratio of
\((W, A)\) at \(y\) equals \(\delta^2/2 - \delta(y - \gamma z)\),
nonzero for \(y \ne \gamma z + \delta/2\); and \(Z\)--\(A \mid W, Y\),
since the log odds ratio of \((Z, A)\) given \((W = w, Y = y)\) equals
\(\mathrm{logit}\, e_1 - \mathrm{logit}\, e_0 - \gamma\theta +
\gamma\delta w\), whose two values differ by \(\gamma\delta \ne 0\). The
remaining five statements --- \(Z \perp Y\), \(A \perp Y\),
\(Z \perp W \mid Y\), \(W \perp A \mid Y\) and \(Z \perp A \mid Y\) ---
can hold at special parameter values, but each holds only where a
real-analytic function of the parameter vanishes: for the first two, the
mean gaps \(\gamma + (\theta - \delta q)(e_1 - e_0)\) and
\(\gamma\{E(Z \mid A=1) - E(Z \mid A=0)\} + \theta - \delta q\); for the
last three, the conditional log odds ratio at a fixed \(y\), say
\(y = 0\), which is real-analytic in the parameter on the connected open
set \((0,1)^4 \times \mathbb{R}^3\) containing the parameter space (an
independence holding for almost every \(y\) forces the log odds ratio to
vanish at every \(y\), by continuity). None of these five functions
vanishes identically --- at the printed instance they equal \(1.058\),
\(0.536\), \(-0.725\), \(1.394\) and \(0.410\) --- so each zero set is
closed, Lebesgue-null and without interior. The non-faithful members of
\(\mathcal{F}\) therefore form a Lebesgue-null set with empty interior;
faithful members exist in each of the open regions
\(\{e_0, e_1 < 1/2\}\) and \(\{e_0, e_1 > 1/2\}\), and the printed
instance and its \(e \mapsto 1-e\) mirror satisfy all twenty dependences
(the five witnesses at the mirror are \(0.942\), \(-0.036\), \(-1.005\),
\(1.740\) and \(-1.987\)). Two remarks. The context-specific
independence \(W \perp Y \mid A = 1\), which is what makes \(W\) a
control-arm-only predictor, is not a violation of faithfulness:
faithfulness concerns \(W \perp Y \mid A\), which fails because the
\(A = 0\) stratum is dependent. And the penalty in Theorem 2(ii) does
not require faithfulness at all, since (4) holds for every member of
\(\mathcal{F}\).

\emph{Closed forms.} Under \(S_2 = \{Z, W\}\):
\(\sigma_1^2 = \sigma_0^2 = 1\), \(e_{S_2} = e(Z)\),
\(\tau_{S_2} = \theta - \delta W\), and by \(W \perp (Z, A)\),
\(\psi = \theta - \delta q\) and
\(E[e (\tau - \psi)^2] = E[e]\,\delta^2 \mathrm{var}(W) = p\, v_w\);
summing the three terms of (1) gives
\(p^2 V_{\mathrm{att}}(S_2) = p + H + p\,v_w\). Under \(S_1 = \{Z\}\):
\(\mu_0(Z) = \gamma Z + \delta q\), so
\(\sigma_0^2(Z) = 1 + \delta^2 \mathrm{var}(W) = 1 + v_w\) (the law of
\(W\) given \(Z, A=0\) is \(\mathrm{Bern}(q)\) because \(e\) is free of
\(W\)), \(\sigma_1^2(Z) = 1\), and
\(\tau_{S_1} = \theta - \delta q = \psi\) identically, so the third term
vanishes: \(p^2 V_{\mathrm{att}}(S_1) = p + H(1 + v_w)\). Subtracting
gives (4); \(p - H = E[e - \kappa(e)] = E[e(1-2e)/(1-e)]\). The ATE
display follows the same way from the ATE bound:
\(V_{\mathrm{ate}}(S_1) = E[1/e] + (1+v_w)E[1/(1-e)]\),
\(V_{\mathrm{ate}}(S_2) = E[1/e] + E[1/(1-e)] + v_w\), difference
\(-v_w E[e/(1-e)] < 0\).

\emph{Theorem 2.} If \(e_0, e_1 < 1/2\) then \(e(1-2e)/(1-e) > 0\)
pointwise, so (4) is strictly positive and \(\{Z\}\) is the unique
minimizer; if \(e_0, e_1 > 1/2\) the sign is reversed and \(\{Z, W\}\)
is the unique minimizer. Claims (i) and (ii) follow; the ATE contrast
follows from the last display of Lemma 3. \(\square\)

\subsection*{A.4 Proof of Theorem 3}\label{a.4-proof-of-theorem-3}
\addcontentsline{toc}{subsection}{A.4 Proof of Theorem 3}

Write \(B = O \setminus Z\) and \(R = Z \setminus O\), so
\(Z \cup O = Z \cup B = O \cup R\). \emph{Step 1 (supplement \(Z\) by
\(B\)).} By Lemma 4(ii), \(A \perp_G B \mid Z\), so \(B \perp A \mid Z\)
under \(P\) and, by Lemma 4(iii), \(Z \cup B\) is valid; Theorem 1(b)
applies with \((S, W) = (Z, B)\). Under no effect modification,
\(\mu_1(Z \cup B) - \mu_0(Z \cup B) = \tau_{Z\cup B}\) is constant, so
Corollary 3 gives \(V_{\mathrm{att}}(Z) - V_{\mathrm{att}}(Z \cup O)
= p^{-2} E[ e_Z (1-e_Z)^{-1} \mathrm{var}\{\mu_0(Z\cup O) \mid Z\} ] \ge 0\).
\emph{Step 2 (delete \(R\) from \(O \cup R\)).} By Lemma 4(ii),
\(Y \perp_G R \mid O, A\), so \(Y \perp R \mid A, O\) under \(P\), and
\(O\) is valid by Lemma 4(i); Theorem 1(a) applies with
\((S, W) = (O, R)\) and gives
\(V_{\mathrm{att}}(Z \cup O) - V_{\mathrm{att}}(O)
= p^{-2} E[\sigma_0^2(O)\{E(\kappa(e_{Z\cup O}) \mid O) - \kappa(e_O)\}] \ge 0\).
Adding the two steps yields (5); the equality conditions are those of
Corollary 3 and Theorem 1(a). \(\square\)

\subsection*{A.5 Proof of Lemma 2 and Theorem
4}\label{a.5-proof-of-lemma-2-and-theorem-4}
\addcontentsline{toc}{subsection}{A.5 Proof of Lemma 2 and Theorem 4}

\emph{Lemma 2.} With the notation of A.1, write \(\psi_h = N/D\),
\(N = E\{h(\eta)\tau_S\}\), \(D = E\{h(\eta)\}\). Along a submodel,
\(\dot N = E[h(\eta)\tau_S g_S] + E[h'(\eta)\tau_S \dot\eta] + E[h(\eta)\dot\tau]\)
and \(\dot D = E[h(\eta) g_S] + E[h'(\eta)\dot\eta]\), so \[
\dot\psi_h = \frac{1}{D}\Bigl( E[h(\eta)(\tau_S - \psi_h) g_S]
 + E[h'(\eta)(\tau_S - \psi_h)\dot\eta] + E[h(\eta)\dot\tau] \Bigr).
\] Matching as in A.1 gives \(\varphi_{h,S}\) as displayed, with
\(\varphi_1 = h(\tau_S - \psi_h)\),
\(\varphi_2 = h'(\tau_S - \psi_h)(A - \eta)\),
\(\varphi_3 = h\{A(Y-\mu_1)/\eta - (1-A)(Y-\mu_0)/(1-\eta)\}\), each in
its component of the tangent space and mutually orthogonal by the same
conditional-mean argument. The second moments are
\(E[h^2(\tau_S-\psi_h)^2]\), \(E[h'^2(\tau_S-\psi_h)^2\eta(1-\eta)]\),
and \(E[h^2\{\sigma_1^2/\eta + \sigma_0^2/(1-\eta)\}]\) (the last
because \(E[A(Y-\mu_1)^2/\eta^2] = E[\sigma_1^2/\eta]\) and likewise for
the control arm, with no cross term); summing and dividing by \(D^2\)
gives (2). For \(h \equiv 1\) and \(h(e) = e\) the displayed reductions
are immediate. \(\square\)

\emph{Theorem 4.} \(W \perp A \mid S\) gives \(e_{S \cup W} = e_S\),
hence \(h(e_{S\cup W}) = h(e_S)\), \(h'(e_{S \cup W}) = h'(e_S)\),
\(D_{S\cup W} = D_S\), and
\(\psi_h(S \cup W) = E\{h(e_S)\,E(\tau_{S\cup W} \mid S)\}/D_S =
\psi_h(S)\) since \(\tau_S = E(\tau_{S\cup W} \mid S)\) by the argument
of A.2(b). The same conditional decompositions give
\(E\{\sigma_a^2(S\cup W) \mid S\} = \sigma_a^2(S) - V_a(S)\) and
\(E\{(\tau_{S\cup W} - \psi_h)^2 \mid S\} = (\tau_S - \psi_h)^2 + V_1 + V_0 - 2C_{10}\).
Substituting into (2) at \(S \cup W\) and at \(S\), with \(h, h', e\)
all \(\sigma(S)\)-measurable, \[
D_S^2 \Delta = E\Bigl[ -\frac{h^2}{e} V_1 - \frac{h^2}{1-e} V_0
 + \{h^2 + h'^2 e(1-e)\}(V_1 + V_0 - 2C_{10}) \Bigr],
\] and collecting the coefficients of \(V_1\), \(V_0\), and \(C_{10}\)
gives (6). \emph{Corollary 7:} the integrand of (6) is
\(V_1 a_1 + V_0 a_0 - 2 C_{10} K = \operatorname{tr}(M\Sigma)\) with
\(a_1 = K - h^2/e = h'^2 e(1-e) - h^2(1-e)/e\),
\(a_0 = K - h^2/(1-e) = h'^2 e(1-e) - h^2 e/(1-e)\), and
\(\det M = a_1 a_0 - K^2 = -K h^2/\{e(1-e)\} + h^4/\{e(1-e)\}
= h^2\{h^2 - K\}/\{e(1-e)\} = -h^2 h'^2\). (a) If \(h' = 0\) then
\(K = h^2\) and \(M = -h^2\, w w^{\top}\) with
\(w = \bigl(\{(1-e)/e\}^{1/2}, \{e/(1-e)\}^{1/2}\bigr)^{\top}\), so
\(\operatorname{tr}(M\Sigma) = -h^2\, w^{\top}\Sigma w \le 0\) for every
\(\Sigma \succeq 0\) --- this is the arithmetic--geometric-mean bound of
Corollary 4 in matrix form --- and integrating gives
\(V_h(S \cup W) \le V_h(S)\) since \(D_{S \cup W} = D_S\). (b) Where
\(h'(e) \ne 0\), \(\det M(e) < 0\), so \(M(e)\) has a positive
eigenvalue \(\lambda_+(e)\) with a unit eigenvector
\(v(e) = (v_1, v_0)\) that can be chosen measurably in \(e\); put
\(v(S) = v(e_S)\) on \(\{h'(e_S) \ne 0\}\) and \(v(S) = 0\) elsewhere.
Start from any law in which \(S\) is valid, with potential outcomes
\(Y^{\circ}(a)\), so that
\((Y^{\circ}(0), Y^{\circ}(1)) \perp A \mid S\). Conditionally on \(S\),
draw \(W \sim\) Bernoulli\((1/2)\) independently of
\((A, Y^{\circ}(0), Y^{\circ}(1))\), and define
\(Y(a) = Y^{\circ}(a) + v_a(S)(2W - 1)\). Then
\((Y^{\circ}(0), Y^{\circ}(1), W) \perp A \mid S\), since the joint
conditional law factorizes as the law of \(W\) given \(S\) times that of
\((Y^{\circ}(0), Y^{\circ}(1))\) given \(S\) times that of \(A\) given
\(S\); hence \((Y(0), Y(1))\), a function of
\((Y^{\circ}(0), Y^{\circ}(1), W, S)\), is independent of \(A\) given
\(S\) and given \((S, W)\), so both \(S\) and \(S \cup W\) are valid,
and \(e_{S \cup W} = e_S\). Because \(W\) is independent of
\(Y^{\circ}(a)\) given \(S\),
\(E\{Y^{\circ}(a) \mid S, W\} = E\{Y^{\circ}(a) \mid S\} = \mu_a(S)\)
(the last equality follows from the validity of \(S\)), so
\(\mu_a(S \cup W) = \mu_a(S) + v_a(S)(2W - 1)\) and
\(\Sigma(S) = v(S) v(S)^{\top}\), as
\(\mathrm{var}(2W - 1 \mid S) = 1\); and
\(D_S^2\{V_h(S \cup W) - V_h(S)\} = E[\lambda_+(e_S)\,\mathbf{1}\{h'(e_S) \ne 0\}] > 0\).
\emph{Corollary 6:} with \(h = e(1-e)\), \(h' = 1 - 2e\): the
treated-arm coefficient is
\((1-2e)^2 e(1-e) - e(1-e)^3 = e(1-e)\{(1-2e)^2 - (1-e)^2\}
= -e^2(1-e)(2-3e)\), and the control-arm coefficient is
\((1-2e)^2 e(1-e) - e^3(1-e) = e(1-e)\{(1-2e)^2 - e^2\} = e(1-e)^2(1-3e)\);
the threshold statements follow from the signs, and the exchange
\(e \mapsto 1-e\) maps one to the other. \(\square\)

\subsection*{A.6 Proof of Proposition
1}\label{a.6-proof-of-proposition-1}
\addcontentsline{toc}{subsection}{A.6 Proof of Proposition 1}

Under \(Y \perp W \mid A, S\), \(\tau_{S\cup W} = \tau_S\) (A.2a), so
\(\psi_h(S \cup W) = E\{h(e_{S\cup W})\tau_S\}/E\{h(e_{S\cup W})\}
= E\{E(h(e_{S\cup W}) \mid S)\,\tau_S\}/D_{S\cup W}\). Subtract
\(\psi_h(S) = E\{h(e_S)\tau_S\}/D_S\); since
\(E\{E(h(e_{S\cup W}) \mid S)\,\psi_h(S)\} = D_{S\cup W}\psi_h(S)\), the
difference equals
\(E[\{E(h(e_{S\cup W}) \mid S) - h(e_S)\}\{\tau_S - \psi_h(S)\}]/D_{S\cup W}\),
which is (7). If \(h\) is affine,
\(E(h(e_{S\cup W}) \mid S) = h(E(e_{S\cup W} \mid S)) = h(e_S)\) by the
tower property and the drift is zero for every law. Conversely, suppose
\(h\) is not affine on the propensity domain \([c, 1-c]\). Then Jensen's
equality fails for some two-point distribution: there are \(x_1 < x_2\)
in \([c, 1-c]\) and \(\lambda \in (0,1)\) with
\(\lambda h(x_1) + (1-\lambda) h(x_2) \ne h(x)\),
\(x = \lambda x_1 + (1-\lambda) x_2\) (a function satisfying the
equality for every such pair and weight is affine). Build a law with two
equally likely strata of \(S\) and a binary \(W\) independent of \(S\)
and of \((Y(0), Y(1))\) given \(S\): in the first stratum
\(e_{S \cup W}\) takes the values \(x_1\) and \(x_2\) with probabilities
\(\lambda\) and \(1 - \lambda\), so \(e_S = x\); in the second
\(e_{S \cup W} = e_S = v\) with \(h(v) > 0\); set \(\tau_S = 1\) in the
first stratum and \(0\) in the second. Then
\(\psi_h(S) = h(x)/\{h(x) + h(v)\}\) and
\(\psi_h(S \cup W) = \bar h/\{\bar h + h(v)\}\) with
\(\bar h = \lambda h(x_1) + (1-\lambda) h(x_2) \ne h(x)\), so the two
targets differ; no convexity or concavity of \(h\) on an interval is
needed. The tilted-covariance form in the text follows from
\(E[g\{\tau_S - \psi_h(S)\}] = E\{h(e_S)\}\,E_{Q_h}[r\{\tau_S - \psi_h(S)\}]\)
and \(D_{S \cup W} = E\{h(e_S)\}\{1 + E_{Q_h}(r)\}\), since
\(E_{Q_h}(\tau_S) = \psi_h(S)\). For \(h = e(1-e)\), concavity gives
\(E(h(e_{S\cup W}) \mid S) \le h(e_S)\) with strict inequality wherever
\(W\) changes \(e\). \(\square\)

\section*{Appendix B: Numerical verification}\label{sec-verification}
\addcontentsline{toc}{section}{Appendix B: Numerical verification}

The identities enumerated in this paragraph --- the closed forms of
Lemma 3, the increments (3), (5) and (6), the deletion inequality of
Theorem 1(a), the comparison of Theorem 3, the thresholds of Corollary 6
and the drift (7) of Proposition 1 --- were checked numerically on
randomly drawn discrete designs using an exact probability-table engine,
and the ATT and ATO bounds were compared with finite-sample Monte Carlo
results (\texttt{analysis/verify\_theorems.R}, \texttt{verify\_thm3.R},
\texttt{verify\_wate.R}; outputs in \texttt{analysis/output/}).
Specifically: the closed forms of Lemma 3 agree with the exact engine at
the instance of Section 5, \((e_0, e_1) = (0.168, 0.401)\), and at two
further instances, with absolute errors below \(10^{-14}\); the
increment identities (3), (5), and (6) hold with absolute errors below
\(10^{-12}\) --- floating-point noise at the scale of the bounds, of
order \(10^{-13}\) in our runs --- across 300 random designs each (for
(6), separately for the ATE, ATT, and ATO weights); the deletion
increment of Theorem 1(a) is nonnegative in all 300 designs; the
comparison of Theorem 3 finds \(O(G)\) optimal among the graphically
valid sets in all 300 no-modification designs; the ATO thresholds of
Corollary 6 change sign between \(e = 0.30\) and \(0.34\) and between
\(0.66\) and \(0.70\); the drift of Proposition 1 is exactly zero for
the ATE and the ATT and matches the closed form (7) for the ATO; and the
saturated AIPW estimators attain \(n \cdot \mathrm{var}/V\) between
0.938 and 1.002 for the ATT and ATO bounds at
\(n = 20{,}000\)--\(40{,}000\) with 400 replications, whose sampling
variability (a relative standard error of about
\(\sqrt{2/399} \approx 7\) percent for a variance ratio) accounts for
the departures from one (\texttt{verify\_theorems.txt},
\texttt{verify\_wate.txt}; no replication is excluded, and no cell was
empty in either treatment arm at these sample sizes).

The random-structure scan draws 60 structures over four binary
covariates (script \texttt{counterexample\_search.R} in
\texttt{analysis/}, output \texttt{counterexample\_results.txt} in
\texttt{analysis/output/}): each ordered pair of covariates receives an
edge with probability 0.3 (lower-triangular, coefficients uniform on
\([0.6, 1.2]\) on the logit scale with intercept \(-0.3\)), each
covariate enters the treatment logit with probability 0.6 (coefficients
uniform on \([0.8, 1.4]\), intercept \(-1.9\)) and the outcome equation
with probability 0.6 (coefficients uniform on \([0.6, 1.4]\)), and each
outcome parent is an effect modifier with probability 0.5 (modifier
coefficients uniform on \([-1.6, 1.2]\), allowing arm-asymmetric
modification); the outcome error is standard normal. All four covariates
are declared as vertices of the graph, so that a covariate with no edges
is a valid addition to any valid set; the valid sets are those
satisfying the adjustment criterion relative to the drawn graph
(enumerated with the \texttt{dagmv} package). Structures with no
treatment parent or no outcome parent (7 of 60) and structures with
fewer than two valid sets (4 of 60) are discarded. The bounds of every
valid set are computed exactly from the structural parameters by
enumerating the \(2^4\) covariate cells, so that ties and penalties are
exact. A structure is counted as a divergence when the two minimizers
differ and the exact ATT penalty for using the ATE-optimal set exceeds 1
percent; structures whose minimizers differ by a smaller penalty are not
counted, but every positive penalty (relative tolerance \(10^{-9}\)) is
listed in the output together with the unthresholded count. At the
reported seed (99) the two counts coincide, the smallest penalty being
2.9 percent. Two further seeds, whose outputs are included in the
archive, show the threshold at work: with seed 123, 53 structures are
retained and the minimizers differ in 5, one with a penalty of 0.35
percent and the other four with penalties of 16.5 to 24.6 percent, so
that 4 are counted; with seed 2026, 56 are retained and the minimizers
differ in 6, one with a penalty of 0.01 percent and the other five with
penalties of 1.5 to 26.9 percent, so that 5 are counted. Exact ties are
common --- in 21 of the 49 retained structures two or more valid sets
attain the ATT or the ATE minimum (relative tolerance \(10^{-9}\)), for
instance because a covariate that predicts neither treatment nor outcome
given a minimizer can be added to it without changing either bound ---
and are handled as follows: every tied set is listed in the output, each
structure is counted once, regardless of which tied set is named, and
the penalty for using the ATE-optimal set is unaffected, because in
every retained structure all sets tied for one minimum have the same
value of the other bound (to within a relative \(10^{-14}\)). An earlier
version of the scan used cell averages on a simulated population of
\(4 \times 10^5\) and an edge-only graph declaration, which excluded
sets containing an edgeless covariate and gave penalties within 0.55
percentage points of the exact ones.

\section*{Reproducibility}\label{reproducibility}
\addcontentsline{toc}{section}{Reproducibility}

The replication archive --- scripts, the two data files, and the outputs
from which every table entry and every number attributed to a script is
transcribed --- is available at
\url{https://github.com/sokubo/paper-estimand-adjustment-replication}
(archive version 1.1; the snapshot checked against this version of the
paper is commit
\href{https://github.com/sokubo/paper-estimand-adjustment-replication/tree/a007c8e7a8c0793784f6c72cbf5e9d7ea3a86cfe}{\texttt{a007c8e}}).
Its \texttt{README.md} maps each table, figure and script-produced
number to a script, an output file and a seed. The entry points are
\texttt{verify\_theorems.R} (seeds 1--3), \texttt{verify\_thm3.R} (seed
7) and \texttt{verify\_wate.R} (seeds 11--12) for the exact-engine
checks of Appendix B, \texttt{counterexample\_search.R} (seed 99 for the
scan; seeds 123 and 2026, set through the environment variable
\texttt{SCAN\_SEED}, for the two supplementary runs of Appendix B) for
the random-structure scan, \texttt{sim\_finite\_sample\_v2.R} (seed
2026) for Table 2, \texttt{apply\_lalonde\_att\_v2.R} (deterministic)
for Table 3 and \texttt{fig\_signmap.R} for Figure 1; the first-version
scripts \texttt{sim\_finite\_sample.R}, \texttt{apply\_lalonde\_att.R}
and \texttt{counterexample\_search\_v1\_mc.R} are retained as legacy
files; none of their outputs is reported here. The shipped outputs were
produced with R 4.6.0, \texttt{dagmv} 0.1.2
(\url{https://github.com/sokubo/dagmv}, used only to enumerate valid
adjustment sets in the scan) and \texttt{ggplot2} 4.0.3 (Figure 1 only);
those of the first arXiv version, produced with R 4.3.3 and
\texttt{ggplot2} 3.4.4, differ from them only at machine precision, and
that version was also reproduced independently under R 4.6.0. All other
computations use base R, and no script runs for more than about a
minute. No new standalone software package is introduced; the analysis
scripts and data accompany the paper. The LaLonde data are the public
NSW (Dehejia--Wahba subsample) and CPS-1 control files of Dehejia and
Wahba (\citeproc{ref-dehejiawahba1999}{1999},
\citeproc{ref-dehejiawahba2002}{2002}), as distributed by Rajeev Dehejia
(\texttt{nsw\_dw.dta} and \texttt{cps\_controls.dta} at
\url{https://users.nber.org/~rdehejia/data/}), read into R and saved
unchanged as \texttt{analysis/data/nsw\_dw.rda} (185 treated and 260
experimental controls; SHA-256 beginning \texttt{6b4637bd}) and
\texttt{analysis/data/cps\_controls.rda} (15,992 controls; SHA-256
beginning \texttt{0c41b180}); the script
\texttt{analysis/data/get\_lalonde\_data.R} verifies the two files
against their row counts and the experimental benchmark and, if they are
absent, rebuilds them from the CRAN package \texttt{causaldata}
(datasets \texttt{nsw\_mixtape} and \texttt{cps\_mixtape}, which we
checked to be value-identical to the shipped files and on which
\texttt{apply\_lalonde\_att\_v2.R} reproduces Table 3 exactly) or from
Dehejia's site. The sample construction and variable definitions of
Section 9 are stated in the archive's \texttt{README.md} and implemented
in \texttt{apply\_lalonde\_att\_v2.R}.

\section*{Acknowledgements}\label{acknowledgements}
\addcontentsline{toc}{section}{Acknowledgements}

I thank Jürgen Degenfellner for reproducing the replication archive of
the first arXiv version independently and for pointing out that the
divergence count of the random-structure scan applies a 1 percent
penalty threshold that the text did not state; the present version
states it and reports the unthresholded count alongside.

\section*{References}\label{references}
\addcontentsline{toc}{section}{References}

\phantomsection\label{refs}

\end{document}